\documentclass[preprint, 5p]{elsarticle}
\usepackage{nth}

\usepackage{mathtools}
  \mathtoolsset{showonlyrefs, centercolon, mathic}
\usepackage{amssymb}
\usepackage{dsfont}
\usepackage{nicefrac}
\usepackage{upgreek}
\usepackage{siunitx}
\usepackage[usenames, dvipsnames, svgnames]{xcolor}
\usepackage{tikz}
  \usetikzlibrary{arrows}
    \tikzset{>=stealth}
  \usetikzlibrary{backgrounds}
  \usetikzlibrary{external}
    \tikzset{external/only named}
    \tikzset{external/mode=list and make}
    \tikzset{external/prefix=}
    \tikzset{external/system call=%
        {pdflatex \tikzexternalcheckshellescape -synctex=1 %
        -interaction=batchmode -jobname "\image" "\texsource"}%
    }
    \pgfkeys{/pgf/images/include external/.code={%
        \IfFileExists{#1.pdf}{%
          \includegraphics{{#1}.pdf}%
        }{%
          \includegraphics[draft]{{#1}.pdf}%
        }%
      }
    }
  \usetikzlibrary{math}
  \usetikzlibrary{plotmarks}
  \usetikzlibrary{positioning}

\usepackage{pgfplots}
  \usepgfplotslibrary{groupplots}
  \usepgfplotslibrary{statistics}
  \pgfplotsset{compat=newest}
  \pgfplotsset{
    scaled ticks=false,
    clip marker paths=true,
    xlabel near ticks,
    ylabel near ticks,
    tick label style={font=\footnotesize},
    label style={font=\footnotesize},
    every axis title shift=1.4pt,
  }

\usepackage{booktabs,colortbl}
\usepackage{pgfplotstable}
  \pgfplotstableset{
    every odd  row/.style={before row={\rowcolor[gray]{0.9}}},
    every head row/.style={before row=\toprule,after row=\midrule},
    every last row/.style={after row=\bottomrule},
  }

\usepackage{etoolbox}
\usepackage{enumerate}
\usepackage{dblfloatfix}
\usepackage{paralist}
\usepackage[multidot]{grffile}
\usepackage{needspace}
\usepackage{xifthen}

\usepackage[hidelinks]{hyperref}

\usepackage{xparse}
\makeatother
\let\OldNabla\nabla
\RenewDocumentCommand{\nabla}{e_}{%
    \OldNabla
    \IfValueT{#1}{%
        _{\!#1}
    }%
}
\makeatletter

\newcommand{\imptrue}{{h_k^{(i,j)}}} % number of parameters
\newcommand{\impest}{{\tilde{h}_k^{(i,j)}}} % number of parameters
\newcommand{\npar}{{n_{\uptheta}}} % number of parameters
\newcommand{\nxi}{{n_{\upxi}}} % number of xi samples
\newcommand{\nnu}{{n_{\upnu}}} % number of nu samples
\newcommand{\nx}{{n_{\mathrm{x}}}} % number of states
\newcommand{\ninp}{{n_{\mathrm{u}}}} % number of states
\newcommand{\ny}{{n_{\mathrm{y}}}} % number of measurements
\newcommand{\zcond}{{\zeta_{\mathrm{cond}}}} % Conditional zeta
\newcommand{\Scross}{{S_{\mathrm{cross}}}} % Cross S
\newcommand{\reals}{{\mathds{R}}} % Real numbers
\newcommand{\dd}{{\mathrm{d}}} % differential d
\newcommand{\opt}{^{*}} % Optimal
\newcommand{\sopt}{^{\star}} % Optimal with star
\newcommand{\trans}{^{\mathsf{T}}} % Transpose
\newcommand{\inv}{^{-1}} % Inverse
\newcommand{\mbj}{^{(j)}} % Mini-batch j
\newcommand{\isamp}{^{(i)}} % Sample i
\newcommand{\elbo}{\mathcal{L}} % Evidence lower bound
\newcommand{\normald}{\mathcal{N}} % Normal distribution
\DeclareMathOperator{\ident}{I} % Identity matrix
\DeclareMathOperator{\const}{const} % constant terms
\DeclareMathOperator{\E}{E} % Expectation
\DeclareMathOperator{\kld}{KL} % Kullback--Leibler divergence
\DeclareMathOperator{\tr}{tr} % Trace of a matrix
\DeclareMathOperator{\entro}{h} % Entropy
\DeclareMathOperator{\vech}{vech} % Vectorize half
\DeclareMathOperator{\matl}{matl} % Matrix from lower half
\DeclareMathOperator{\tria}{tria} % Matrix triangularization
\DeclareMathOperator*{\argmax}{arg\,max}

\newcommand{\lstdw}{{\varsigma_{\operatorname w}}}
\newcommand{\lstdv}{{\varsigma_{\operatorname v}}}
\newcommand{\tfin}{{t_{\operatorname f}}}
\newcommand{\Tsim}{{\Delta t_{\operatorname {sim}}}}
\newcommand{\Ts}{{T_{\operatorname s}}}

\journal{Automatica}
\biboptions{authoryear}

\begin{document}

\begin{frontmatter}
  \title{%
    Parameterizations for Large-Scale Variational System Identification Using Unconstrained Optimization\tnoteref{t1}%
  }
  \tnotetext[t1]{%
    This research was made possible by the NASA Established Program to Stimulate Competitive Research, Grant \#80NSSC22M0027.
    Computational resources were provided by the WVU Research Computing Thorny Flat HPC cluster, which is funded in part by NSF OAC-1726534.
  }

  \author{Dimas Abreu Archanjo Dutra\corref{cor}}
  \ead{dimas.dutra@mail.wvu.edu}
  %\cortext[cor]{Corresponding author}
  
  \address{%
    Department of Mechanical and Aerospace Engineering, Statler College of Engineering and Mineral Resources, West Virginia University, Morgantown, WV, USA
  }
    
  % Abstract should have no more than 200 words.
  \begin{abstract}
    This paper details how to parameterize the posterior distribution of state-space systems to generate improved optimization problems for system identification using variational inference.
    Three different parameterizations of the assumed state-path posterior distribution are proposed based on this representation: time-varying, steady-state, and convolution smoother; each resulting in  a different parameter estimator.
    In contrast to existing methods for variational system identification, the proposed estimators can be implemented with unconstrained optimization methods.
    Furthermore, when applied to mini-batches in conjunction with stochastic optimization, the convolution-smoother formulation enables identification of large linear and nonlinear state-space systems from very large datasets.
    For linear systems, the method achieves the same performance as the inherently sequential prediction-error methods using an embarrassingly parallel algorithm that benefits from large speedups when computed in modern graphical processing units (GPUs).
    The ability of the proposed estimators to identify large models, work with large datasets split into mini-batches, and work in parallel on GPUs make them well-suited for identifying deep models for  applications in systems and control.
  \end{abstract}

\end{frontmatter}

%\runningpagewiselinenumbers
%\linenumbers

\section{Introduction}
Bayesian statistics is the framework of choice for representing uncertainty in dynamical systems \citep{sarkka_bayesian_2013} and signal processing \citep{tzikas_variational_2008, candy_bayesian_2016}, but it leads to \emph{intractable} inference problems in all but a limited class of cases, like those with linear--Gaussian systems.
In general, the posterior lacks a closed-form expression, making the calculation of moments or sampling very difficult, but the complete data probability density and its derivatives can be easily evaluated pointwise, enabling maximum \emph{a posteriori} (MAP) estimation.

With advances in computing power and nonlinear optimization tools, MAP estimation has emerged as a a tractable generalization of Kalman filtering and smoothing \citep{aravkin_optimization_2014, aravkin_generalized_2017}, as it coincides with these classic algorithms in the linear--Gaussian case but is applicable to a wider class of nonlinear and non-Gaussian estimation problems.
In particular, as it is applicable to problems with heavy-tailed distributions, MAP can be used for robust estimation in the presence of outliers \citep{aravkin_optimization_2014, aravkin_generalized_2017, dutra_maximum_2014, dutra_joint_2017, farahmand_doubly_2011}.
Nevertheless, the reason for which MAP estimation is tractable is also its main limitation: it builds a \emph{global} approximation to the posterior using its \emph{local} properties at a single point.
This is true even when uncertainty estimates based on Laplace's method are used \citep{dutra_uncertainty_2020}, as they are built from local properties at the mode \citep[Sec.~4.1]{zhang_advances_2019}.

Variational inference (VI) overcomes these limitations by building a global approximation of the posterior which minimizes the mismatch between the true and approximate density across the whole sample space \citep{tzikas_variational_2008, zhang_advances_2019}.
For parameter estimation, VI yields a lower bound on the true, intractable, likelihood function.
A series of recent publications introduced a formulation of VI for nonlinear state-space systems that matches the state-of-the-art performance in state estimation \citep{courts_gaussian_2021}
and system identification \citep{courts_variational_2021, courts_variational_2023} at a lower computational cost and significantly improved convergence rates.

The primary contribution of this paper is proposing three novel parameterizations of the posterior distribution of Gaussian state-space systems for implementing variational system identification \citep{courts_variational_2021, courts_variational_2023} with unconstrained optimization.
Using matrix triangularization, like in square-root Kalman filtering \citep[Chap.~12]{kailath_linear_2000}, the proposed parameterizations lead to smaller optimization problems with objective functions that are embarrassingly parallelizable, benefiting from large speedups when executed in graphical processing units (GPUs).
This representation also leads, naturally, to steady-state estimators much like those used in prediction-error methods for linear \citep{maine_formulation_1981, wills_gradient-based_2008} and nonlinear systems \citep[Chap.~5]{jategaonkar_flight_2015}.
Approximating the steady-state estimator using a convolution smoother, furthermore, enables the number of decision variables of the optimization to be fixed, independent of the dataset size, enabling the use of stochastic optimization \citep{kingma_adam_2015, schmidt_descending_2021, wills_stochastic_2021} to perform variational system identification in very large datasets.
Additional contributions of the paper are listed in the end of Sec.~\ref{sec:relw}.

\section{Related Work}
\label{sec:relw}

A major difficulty of system identification of state-space models is that the states---a defining property of the model---are not directly observable \citep[Sec.~1]{courts_variational_2023}.
Therefore, many parameter estimators for this class of models employ an auxiliary state estimator to circumvent this limitation.
The class of maximum likelihood estimators, to which our method belongs, can be broadly divided into prediction-error, joint maximum \emph{a posteriori} (MAP), and expectation--maximization (EM) methods.

Prediction-error methods \citep[Sec.~7.2]{astrom_maximum_1980, ljung_system_1999} employ a state-space filter to estimate the one-step ahead predictive distribution, which is then used to compute the likelihood function and its derivatives, for optimization.
For linear--Gaussian systems, the filter---and, consequently, the likelihood function---has a known, tractable solution: the Kalman filter.
For nonlinear or non-Gaussian systems, approximate filters such as the extended Kalman filter \citep{kristensen_parameter_2004}, sigma-point Kalman filters \citep{kokkala_sigma-point_2016}, or particle filters \citep[Sec.~IV-B-1]{andrieu_particle_2004} are used instead.

Even when the filter--predictor has an analytical solution, however, a simpler one might be employed instead.
In the aircraft system identification literature, for example, steady-state Kalman filters \citep{maine_formulation_1981, grauer_new_2015} and extended Kalman filters \citep[Chap.~5]{jategaonkar_flight_2015} are used instead of their time-varying forms.
The problem of estimating a state-space model is also frequently reframed as the estimation of a linear predictor \citep{mckelvey_data_2004, wills_gradient-based_2008}.
While the steady-state Kalman filter is a linear predictor, not all linear predictors estimated in this form will be associated with a Kalman filter (see, e.g., discussion by \citealp[Sec.~4]{maine_formulation_1981}).

Joint MAP approaches are based on constructing a Gaussian approximation of the smoothed state posterior at the mode.
If a Bayesian treatment of the parameters is used, the joint mode of the state-path and parameters can be obtained \citep{dutra_joint_2017}.
Better parameter estimates are obtained, however, by marginalizing the states using Laplace's method \citep{karimi_approximate_2013, karimi_maximum-likelihood_2014, karimi_bayesian_2018} to approximate the likelihood function.

Expectation--maximization methods employ a smoother to estimate all states, which are in turn used to maximize a lower bound of the likelihood function.
Once again, for linear--Gaussian systems the smoother has an analytical form, the Kalman smoother, as used by \citet{gibson_robust_2005}.
For nonlinear systems, an approximation such as a particle smoother \citep{schon_system_2011} or a sigma-point smoother \citep{kokkala_sigma-point_2016} must be used.

One notable limitation of prediction-error and expectation maximization approaches for maximum likelihood estimation, however, is that simplifications and approximations in the state estimator cannot be decoupled from the state-space model estimated.
In prediction-error methods, for example, the model defines the predictor, so different predictors for the same model cannot be compared.
This had led to debate on whether improved approximations of integrals are beneficial \citep[Sec. V]{kokkala_sigma-point_2016} or not \citep{arasaratnam_cubature_2009} when applying Gaussian estimators to systems with non-Gaussian posteriors.
This is a stark contrast to variational inference \citep{courts_variational_2021, courts_variational_2023}, where the state-space model and the approximation of the smoothed states can be chosen idependently.
A simplified estimator that yields a state posterior farther from the true one, as quantified by the Kullback--Leibler divergence, will add more slack to the lower bound of the likelihood function computed by VI.

In this paper, we bring together concepts from all these approaches to variational system identification.
We frame VI as a generalization of joint MAP estimation with Laplace's method for marginalizing the states.
We show that the parameterization of the state posterior proposed by \citet{courts_variational_2021, courts_gaussian_2021, courts_variational_2023} represents a Bayesian network \citep[Sec.~1.3]{dellaert_factor_2017}.
Additionally, we show that steady-state formulations and simplified state estimators, like used in prediction-error methods, can make the problem scalable to large datasets and models.
Like in expectation--maximization approaches, these simplified estimators are smoothers, but different estimators can be plugged into the same model, for system identification, and compared.

\section{Background}
\subsection{Notation and conventions}
\label{sec:notation}
Throughout this paper, sequences $x_0,\dotsc,x_N$ will be represented with the shorthand notation $x_{0:N}$.
Random variables will be written in uppercase letters and values they might take will be written in lowercase.
Expectations with respect to a density $q\colon\reals^n\to\reals$ will be denoted by $ \E_{X\sim q}[f(X)]  = \E_q[f(X)] := \int_{\reals^n} f(x)q(x)\,\dd x$.
Lower-triangular matrices will be converted to vectors using the $\vech$ operator, as defined by \citet[Sec.~3.8]{magnus_matrix_2019}, which stacks the elements of a matrix into a vector, column by column, excluding all entries above the main diagonal.
We also define a $\matl$ operator as the inverse of $\vech$ for square lower triangular matrices, i.e., $L = \matl(\vech(L))$.
A general triangularization routine of $M\in\reals^{n\times m}$, with $n\leq m$, will be denoted by $S=\tria(M)$, resulting in a lower triangular $S\in\reals^{n\times n}$ such that $SS\trans = MM\trans$.
It can be computed, e.g., with the $R\trans$ factor of the reduced QR decomposition of $M\trans$ \citep[Chap.~12]{kailath_linear_2000}.

\subsection{Variational Inference}
In this section, we derive the VI objective function in a way that frames its maximization as a generalization of MAP estimation.
Our approach is complementary to that of \citet[Sec.~3.1]{courts_variational_2023} which framed VI as a generalization of expectation--maximization.

From Bayes' rule, we have that the model evidence, or likelihood of a vector $\theta\in\reals^\npar$ of unknown parameters, is the ratio between the density of the complete data and the posterior
\begin{equation}
  \label{eq:bayes}
  p_\theta(y) = \frac{p_\theta (x,y)}{p_\theta(x|y)}
\end{equation}
at every point in the support of the posterior, where $y$ is the vector of observations and $x$ the vector of latent variables.
This expression is not practical to compute the likelihood function, however, since the computation of the posterior density is usually intractable or too complex to use in an optimization problem.

Eq.~\eqref{eq:bayes} can, nevertheless, be used to build a lower bound to the likelihood function if an approximation $q(x)$ to the posterior density, called the assumed density, is introduced.
As long as the support of $q$ is contained in the support of the posterior%
\footnote{%
  Or, equivalently, the measure associated with $q$ is absolutely continuous with respect to the posterior probability measure.
},
\begin{equation}
  \label{eq:loglike_with_q}
  \log p_\theta(y) = 
  \log\left(\frac{p_\theta (x,y)}{q(x)}\right)
  + \log\left(\frac{q(x)}{p_\theta(x|y)}\right).
\end{equation}
The first term of the right-hand side of \eqref{eq:loglike_with_q} can be computed, as it requires only the complete data density which defines the model and the assumed density, which is chosen.
The second term, on the other hand, is close to zero whenever $q$ is close to the posterior.
By taking the expectation of both sides with respect to $q$, we have that the log-likelihood is equal to the sum of a computationally tractable expression and an indicator of the distance between the approximation and the posterior throughout the whole sample space:
\mathtoolsset{showonlyrefs=false}%
\begin{align}
  \label{eq:vi_bound}
  \log p_\theta(y) &= 
  \underbrace{
    \E_q\!\left[\log\left(\frac{p_\theta (X,y)}{q(X)}\right)\right]
  }_{\mathclap{\elbo(\theta, q)}}
  + 
  \underbrace{
    \E_q\!\left[\log\left(\frac{q(X)}{p_\theta(X|y)}\right)\right]
  }_{\mathclap{\kld[q(x)\, \Vert\, p(x|y)]\geq 0}},
  \\
  \notag
  \log p_\theta(y) &\geq \elbo(\theta, q).
\end{align}
\mathtoolsset{showonlyrefs}

The first term on the right-hand side of \eqref{eq:vi_bound} is the variational objective $\elbo(\theta, q)$, while the second term is the Kullback--Leibler divergence (KLD) of $q$ from the posterior.
The KLD, also known as relative entropy, quantifies the difference between two probability distributions and is non-negative, being zero if and only if both densites are equal almost everywhere \citep[see, e.g.,][Sec.~2]{zhang_advances_2019}.
Consequently, the variational objective is a lower bound on the likelihood function, or evidence, for which it is also denoted the evidence lower bound (ELBO).

Joint state and parameter estimation can then be performed by searching the $(\theta, q)$ pair that maximizes the ELBO, which at the same time raises the lower bound on the likelihood and shrinks the approximation error, as quantified by the KLD.
The posterior approximation must be chosen from a space of distributions with respect to which the expectation is tractable to compute.
A final simplification of the ELBO, from the expression in \eqref{eq:vi_bound}, replaces the expected log-density of $X$ with its differential entropy $\entro_q[X]$ under $q$:
\begin{align}
  \label{eq:elbo_entro}
  \elbo(\theta, q) &=
  \E_q\!\left[\log p_\theta (X,y)\right] + 
  \underbrace{
    \E_q\!\left[-\log q(X)\right]
  }_{\entro_q[X]}.
\end{align}

Eq.~\eqref{eq:elbo_entro} also shows that, with respect to the latent variables, VI is a regularized generalization of MAP estimation.
Instead of choosing the single latent variable vector $x\sopt=\argmax_{x} \log p_\theta(x,y)$ which maximizes the complete-data log-density, in VI we choose the population of points $X\sim q$ which maximizes the mean complete-data log-density.
If the posterior has a strict global maximum, however, the population would colapse to a single point at the mode, with $q$ being Dirac's delta, a singular distribution.
Adding the entropy of $q$ as a regularizer, however, prevents that.
The choice of entropy as the regularizer is justified by the fact the optimum of the regularized problem is the closest to the true posterior, as quantified by the KLD.

\subsection{Gaussian Variational Inference}
When the approximation $q$ is a multivariate normal distribution, VI can also be interpreted as a generalization of Laplace's method.
The use of Gaussian $q$ in VI, denoted Gaussian Variational Inference (GVI), is popular due to its simplicity, ease of integration, and the fact that the distribution arises frequently in many problems, as explained by the central limit and Bernsten--von Mises theorems.

Throughout this section, let $q_{\mu, S}(x)=\normald(x; \mu, SS\trans)$ denote the density of a normal approximation with mean $\mu$ and covariance $SS\trans$.
For this approximation, the ELBO reduces to
\begin{equation}
  \label{eq:gvi_elbo}
  \elbo(\theta, q_{\mu, S}) = 
  \E_{Z\sim\normald(0,\ident)}[\log p_\theta(\mu +SZ, y)] + \log|\det S|
  +\const,
\end{equation}
where $Z$ is a vector of independent standard normal variables and ``$\const$'' denotes constant terms which do not change the location of maxima.
In this special case, an important result is that the optimal scaling $S\opt$ satisfies \citep[Eq.~5]{opper_variational_2009}
\begin{equation}
  \label{eq:gvi_opt_cov}
  S\opt (S\opt)\trans = 
  -\left(
    \E_{Z\sim\normald(0,\ident)}[\nabla_{xx}\log p_\theta(\mu +S\opt Z, y)]
  \right)\inv,
\end{equation}
where $\nabla_{xx}$ denotes the Hessian matrix with respect to $x$.
This can be seen as generalization of Laplace's method 
(\citealp[Sec.~2]{opper_variational_2009}; \citealp[Sec.~3.4]{barfoot_exactly_2020}), 
for which the covariance $P\sopt$ is given by inverse of the log-density Hessian at the mode $x\sopt$:
\begin{equation}
  \label{eq:laplace_cov}
  P\sopt = -\left(\nabla_{xx}\log p_\theta(x\sopt, y)\right)\inv.
\end{equation}

Instead of holding pointwise, the GVI approximation of the covariance holds, in average, over the whole sample space, weighted by the best approximation $q_{\mu,S}$ of the posterior density.
For parameter estimation, the presence of the log-determinant of $S$ in  \eqref{eq:gvi_elbo} is equivalent to marginalization of the states with the posterior approximated by Laplace's method
\citep{karimi_approximate_2013, karimi_maximum-likelihood_2014, karimi_bayesian_2018}.
Once again, the key difference is that, in GVI, the approximation is \emph{global}, not local.
In addition, since $S$ is a decision variable of the optimization problem, it can be chosen in a format such that its determinant is trivial to compute, which is usually not the case for Laplace's method.

\section{Problem statement}
\begin{figure}[b]
  \centering
  \tikzsetnextfilename{complete-net.tikz}%
  \tikzpicturedependsonfile{complete-net.tikz.tex}%
  \begin{tikzpicture}[bayesnet/.style={circle, draw, inner sep=0.5mm}]
  \node[bayesnet] (x0) {$x_0$};

  \foreach \i in {1,...,4} {
    \tikzmath{\iprev = \i - 1;}
    \draw [->] (x\iprev) -- ++(0.5, 0)
      node[bayesnet, anchor=west] (x\i) {$x_{\i}$};
  }

  \foreach \i in {0,...,4} {
    \draw [->] (x\i) -- ++(0, 0.5)
      node[bayesnet, anchor=south] (y\i) {$y_{\i}$};
  }

  \draw [->] (x4) -- ++(0.5, 0)
    node[anchor=west] {...};
\end{tikzpicture}%

  \caption{Bayesian network of the complete data, $x_{0:N}$ and $y_{0:N}$.}
  \label{fig:complete-net}
\end{figure}
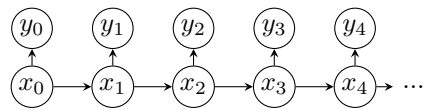
\begin{figure}[b]
  \centering
  \tikzsetnextfilename{posterior-net.tikz}%
  \tikzpicturedependsonfile{posterior-net.tikz.tex}%
  \begin{tikzpicture}[bayesnet/.style={circle, draw, inner sep=0.5mm}]
  \node[bayesnet] (x0) {$x_0$};

  \foreach \i in {1,...,4} {
    \tikzmath{\iprev = \i - 1;}
    \draw [->] (x\iprev) -- ++(0.5, 0)
      node[bayesnet, anchor=west] (x\i) {$x_{\i}$};
  }

  \draw [->] (x4) -- ++(0.5, 0)
    node[anchor=west] {...};
\end{tikzpicture}%

  \caption{Bayesian network of the posterior of the states $x_{0:N}$, given the measurements $y_{0:N}$.}
  \label{fig:posterior-net}
\end{figure}
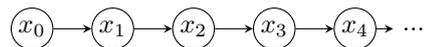
\begin{figure}[b]
  \centering
  \tikzsetnextfilename{factor-graph.tikz}%
  \tikzpicturedependsonfile{factor-graph.tikz.tex}%
  \begin{tikzpicture}[
    bayesnet/.style={circle, draw, inner sep=0.5mm},
    factor/.style={rectangle, fill=black, inner sep=1mm},
  ]
  \node[bayesnet] (x0) {$x_0$};

  \foreach \i in {1,...,4} {
    \tikzmath{\iprev = \i - 1;}
    \draw [-] (x\iprev) -- ++(0.5, 0)
      node[bayesnet, anchor=west] (x\i) {$x_{\i}$}
      node[midway, factor] {};
  }

  \foreach \i in {0,...,4} {
    \draw [-] (x\i) -- ++(0, 0.5)
      node[factor, anchor=south] {};
  }

  \draw [-] (x4.east) -- ++(0.5, 0)
    node[anchor=west] {...}
    node[midway, factor] {};
\end{tikzpicture}%

  \caption{Factor graph of the posterior of the states $x_{0:N}$, given the measurements $y_{0:N}$, with squares representing factors.}
  \label{fig:factor-graph}
\end{figure}
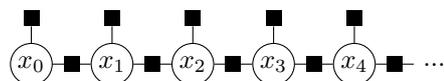

The system identification problem consists of estimating a vector of unknown parameters $\theta\in\reals^\npar$ from a sequence of observations $y_{0:N}$ that are modeled to be generated from the state-space process $x_{0:N}$.
The model is specified by the initial state prior density $p_\theta(x_0)$, transition densities $p_\theta(x_k|x_{k-1})$, and the measurement likelihood $p_\theta(y_{k}|x_k)$.
We assume that the state sequence is a Markov process and that each $y_k\in\reals^\ny$ is conditionally independent, given $x_k\in\reals^\nx$, of the states $x_i$ and measurements $y_i$ at all other time samples $i\neq k$.
We also note that the densities can depend on exogenous inputs $u_k\in\reals^\ninp$, which have been ommited for brevity.

The application the chain rule of probability and the conditional independence properties above lead to the following expression for the joint density of $x$ and $y$, i.e., the complete-data density:
\begin{equation}
  \label{eq:complete-data_dens}
  p_\theta(x_{0:N},y_{0:N}) = 
  p_\theta(x_0)\prod_{k=1}^N p_\theta(x_k|x_{k-1})
  \prod_{k=0}^N p_\theta(y_k|x_k),
\end{equation}
which is the starting point for VI, as seen in \eqref{eq:elbo_entro} and~\eqref{eq:gvi_elbo}.
Another important property of the model, as specified, is that both the complete data and the state posterior can be represented as simple Bayesian networks \citep[Sec.~1.3]{dellaert_factor_2017}, with the topologies shown in Figs.~\ref{fig:complete-net} and~\ref{fig:posterior-net}, respectively.
The network of the complete data follows directly from the problem statement.
To obtain the topology of the state posterior, we first convert the complete data to a factor graph, shown in Fig.~\ref{fig:factor-graph}.
Since the measurement factors are each only connected to a single state, by applying the variable elimination algorithm of \citet{dellaert_factor_2017}, with the states ordered in reverse chronological order, we obtain the Bayesian network of the posterior, 
\begin{equation}
  \label{eq:post_network}
  p_\theta(x_{0:N}|y_{0:N}) = 
  p_\theta(x_{0}|y_{0:N})  
  \prod_{k=1}^N p_\theta(x_{k}|x_{k-1}, y_{0:N}).
\end{equation}

\section{Parameterizations of the assumed density}
\label{sec:param}

In this paper, we propose three different parameterizations for the assumed density with varying degrees of generality.
We start with the most general one, which leads to an unconstrained implementation the problem of \citet{courts_variational_2023}.
In the sequence we show how, for large datasets, the covariances can be represented in steady-state and, finally, how a convolution smoother can be used to obtain the means directly from the data.

In all of the formulations we present, we will restrict the assumed density $q$ to a Bayesian network 
\begin{equation}
  \label{eq:q_network}
  q(x_{0:N}) = q(x_0)\prod_{k=1}^N q(x_k|x_{k-1}),
\end{equation}
of the same topology as the true posterior, which factors as shown in \eqref{eq:post_network} and represented in Fig.~\ref{fig:posterior-net}.
This is a \emph{sparse} representation but does not introduce any approximation error.
We note that this is the same topology used by \citet{courts_gaussian_2021}, without requiring any assumptions beyond those implied by the state-space model\footnote{%
  That is, Assumption~1 of \citet{courts_gaussian_2021} follows directly from the graphical model used.
}.

We further restrict $q$ to the multivariate normal distribution.
Note that, as shown by \citet[Sec.~4]{barfoot_exactly_2020}, Eq.~\eqref{eq:gvi_opt_cov} implies that the optimal approximation of the latent variable information matrix, using GVI, will have the same sparsity structure as the factor graph of the complete-data density.
This means that the optimal multivariate normal $q$ will also factor as \eqref{eq:q_network}.

\subsection{Time-varying parameterization}
\label{sec:time-varying}
To parameterize $q$, it suffices to represent the mean $\mu_{0:N}$ of each state; the covariance of the initial state $\Sigma_0$; the conditional covariance $\{\Sigma_{k|k-1}\}_{k=1}^N$ of each of the following states $x_k$, given the previous;
and the covariance matrices $\{\Sigma_{k, k-1}\}_{k=1}^N$ between $x_k$ and $x_{k-1}$.
From these variables, the marginal covariance $\Sigma_{k}$ of each state can be computed with the following recursion:
\begin{equation}
  \label{eq:marg_squares}
  \Sigma_{k} = \Sigma_{k|k-1} + \Sigma_{k,k-1} \Sigma_{k-1}\inv \Sigma_{k,k-1}\trans,
\end{equation}
which comes from the conditioning of the multivariate normal \citep[see, e.g.,][Sec.~3.C]{kailath_linear_2000}.
Each of the densities in \eqref{eq:q_network} is then given by
\begin{subequations}
  \label{eq:node_densities}
  \begin{align}
    \label{eq:q0}
    q(x_0) &= \normald(x_0; \mu_0, \Sigma_0),
    \\
    \label{eq:qk}
    q(x_k|x_{k-1}) &= 
    \normald\big(
      x_k; \mu_k+\Sigma_{k,k-1}\Sigma_{k-1}\inv(x_{k-1}-\mu_{k-1}),
      \Sigma_{k|k-1}
    \big).
  \end{align}
\end{subequations}\noeqref{eq:q0,eq:qk}%
The parameterization must also ensure the covariances $\Sigma_0$ and $\Sigma_{k|k-1}$ are symmetric positive definite.
To enforce this without constraints, we employ a log-parameterization of the Cholesky factors of the covariances, which are reconstructed using the matrix exponential:
\mathtoolsset{showonlyrefs=false}%
\begin{subequations}
  \label{eq:cov_parameteriz}
  \begin{align}
    S_0 &:= \exp \matl \zeta_0, &
    \Sigma_0 &= S_0 S_0\trans, \\
    \label{eq:cov_param_k}
    S_{k|k-1} &= \exp\matl \zeta_{k|k-1}, &
    \Sigma_{k|k-1} &= S_{k|k-1} S_{k|k-1}\trans,
  \end{align}
\end{subequations}
\mathtoolsset{showonlyrefs}%
where $\zeta_0\in\reals^m$ and $\zeta_{k|k-1}\in\reals^m$ are the vectors with $m=\tfrac 12\nx(\nx+1)$ elements, corresponding to the lower triangles of the matrices.
This approach is similar to that of \citet{glasmachers_exponential_2010}.
The matrix exponential of lower triangular matrices is lower triangular, with algorithms available for its efficient computation \citep{al-mohy_new_2010}.
Equivalent results, however, can also be achieved by  only taking the exponential of the diagonal entries.
The use of log-parameterizations to represent positive quantities---or matrices with positive eigenvalues---is common practice in statistical optimization.

Finally, instead of representing the covariance $\Sigma_{k,k-1}$ directly, we use the Cholesky factors $S_{k}$ of the marginal covariances of \eqref{eq:marg_squares} to define
\begin{equation}
  \label{eq:S_cross_defn}
  S_{k,k-1} := \Sigma_{k,k-1} (S_{k-1}\trans)\inv.
\end{equation}
This approach is reminiscent of square-root filtering \citep[Chap.~12]{kailath_linear_2000} and enables marginalization applying the triangularization routine defined in Sec.~\ref{sec:notation} to the block matrix
\begin{align}
  \label{eq:marg_sqrt}
  S_k &= \tria 
  \begin{bmatrix}
    S_{k|k-1} & S_{k, k-1}
  \end{bmatrix},
  &
  \Sigma_k &= S_k S_k\trans,
\end{align}
which is equivalent to \eqref{eq:marg_squares} using the definition in \eqref{eq:S_cross_defn}.

To summarize, all nonsingular Gaussian $q$ with the Bayesian network topology of \eqref{eq:q_network} can be uniquely represented with the parameters $\zeta_0$, $\{\zeta_{k|k-1}, S_{k,k-1}\}_{k=1}^N$, and $\mu_{0:N}$.
The Cholesky factors\footnote{%
  Note that $S_k$, $S_{k+1, k}$ and $S_{k+1|k}$ are equal to $A_k\trans$, $B_k\trans$ and $C_k\trans$ of \citet{courts_variational_2023}, Eq.~(31).
}
and covariances can be computed with \eqref{eq:cov_parameteriz} and \eqref{eq:marg_sqrt}, yielding all the parameters of the densities in \eqref{eq:node_densities}.
The decision variables of the resulting variational system identification problem are summarized in Tab.~\ref{tab:decision}.

\begin{table}
  \centering
  \pgfplotstableread[col sep=&]{
    Parameterization & Decision Variables
    time varying & $\theta,\mu_{0:N}, \zeta_0, \{\zeta_{k|k-1}, S_{k,k-1}\}_{k=1}^N$
    steady-state & $\theta,\mu_{0:N}, \zcond, \Scross$
    convolution smoother & $\theta,K,\zcond, \Scross$
  }\decvartbl
  \pgfplotstabletypeset[string type, columns/Decision Variables/.style={column type=l}]{\decvartbl}
  \caption{Decision variables of the variational system identification problem for each parameterization of the posterior distribution.}
  \label{tab:decision}
\end{table}

\subsection{Steady-state parameterization}
\label{sec:steady-state}
Under some conditions, the state covariances of both linear and nonlinear systems converge to steady-state values as the dataset grows.
The dynamic nature of state process means that very little information about a given state is added by the knowledge of measurements that are too distant in time.
This fact can be exploited by simplifying the time-varying parameterization of Sec.~\ref{sec:time-varying} to use the same $\zeta_{k|k-1}=\zcond$ and $S_{k,k-1}=\Scross$ for all $k$, leading to a steady-state assumed density $q$, parameterized by $\mu_{0:N}$, $\zcond$, and $\Scross$.
The covariances and Cholesky factors of the states at all instants, including $k=0$, are given by \eqref{eq:cov_param_k} and~\eqref{eq:marg_sqrt}.
The decision variables of the resulting variational system identification problem are summarized in Tab.~\ref{tab:decision}.

When the posterior covariances converge to steady-state values, the difference between the true posterior and the steady-state $q$ will be concentrated at the dataset boundaries, which is not significant as the dataset grows.
This means that the difference between the maximum ELBO achieved with the time-varying and steady-state assumed densities should vanish, if normalized by $N$, as $N\to\infty$.

Note also that, even for linear--Gaussian systems, the covariances and means of the optimal steady-state $q$ will not coincide with those obtained by solving the Riccati equations and applying a steady-state Kalman smoother.
They will correspond to the steady-state $q$ which is closest to the true posterior, as quantified by the KLD.
This will cause the GVI covariances to be slightly higher than the steady-state values, with the difference decreasing as the number of samples increase.
It also means that the method can even be applied to systems that do not reach a steady-state.
Although, in that case, the bound on the likelihood may not be very tight, it is beneficial for the optimization to be able to explore the parameter space without crossing singularities in the objective function.

\subsection{Convolution-smoother parameterization of the mean}
For a given $\theta$, the true posterior can be written as a function of the measurements $y$ and exogenous inputs $u$.
Consequently, the mean $\mu_{0:N}$ can be written as a function of $y_{0:N}$ and $u_{0:N}$.
Once again, due to the dynamic nature of the system, we expect that each $\mu_k$ depends very little on data $y_j,u_j$ that is distant in time, with $|k-j|>M$ greater than a chosen window length $M$.
A simple strategy to limit the growth of the number of decision variables, as $N$ increases, is to represent the mean of the assumed density $q$ as a parameterized function $\phi$ of the data:
\begin{equation}
  \label{eq:smoother_fcn}
  \mu_k = \phi(y_{k-M:k+M}, u_{k-M:k+M}, K),\qquad
  \forall M\geq k\geq N-M,
\end{equation}
where $K$ are the parameters of the function $\phi$ and, indirectly, of the assumed density.

This approach amounts to estimating a smoother for defining the parameters of $q$.
For linear--Gaussian systems, or nonlinear systems operated around an equilibrium condition, a natural choice for $\phi$ is a convolution smoother:
\begin{equation}
  \label{eq:lin_smoother}
  \mu_k = \sum_{j=-M}^{M} K^{(j)}
  \begin{bmatrix}
    y_{k+j} \\ u_{k+j}
  \end{bmatrix},
  \qquad
  \forall M\geq k\geq N-M,
\end{equation}
where $K\in\reals^{\nx\times(\ny+\ninp)\times (2M+1)}$ is a convolution kernel and $K^{(j)}$ indicates indexing at the third dimension, yielding a matrix.
The dataset boundaries can be treated using the same techniques as in convolution: zero-padding, extrapolation, or using a slightly larger dataset for computing the mean than for defining the complete-data density in \eqref{eq:complete-data_dens}.
In large datasets, all these solutions should yield comparable results.

The assumed density $q$ is then parameterized by $K$, $\zcond$, and $\Scross$, as summarized in Tab.~\ref{tab:decision}.
When $N\gg M$, the resulting optimization problem is much smaller than that of Secs.~\ref{sec:time-varying} and~\ref{sec:steady-state}.
Most notably, the number of parameters of the assumed density is fixed, independent of the length of the dataset.

\section{Objective function}
\label{sec:obj}
The parameterizations of Sec.~\ref{sec:param} are not only sparse representations of the posterior which exploit the problem structure, they enable efficient computation of the objective function and its derivatives.
To begin, the chain rule of differential entropy \citep[see, e.g.,][Sec.~8.1]{mackay_information_2019} implies that
\begin{equation}
  \label{eq:chain_rule_entro}
  \textstyle
  \entro_q[X_{0:N}] = \entro_q[X_0] + \sum_{k=1}^N \entro_q[X_k|X_{k-1}],
\end{equation}
where $\entro_q[A|B]$ denotes the conditional entropy of $A$ given $B$, under the distribution $q$.
Substituting the entropy of the multivariate normal,
\begin{align}
  \entro_q[X_{0:N}] 
  &= \textstyle
  \log\det S_0 + \sum_{k=1}^N \log\det S_{k|k-1} + \const
  \\
  \label{eq:entro_norm}
  &= \textstyle
  \tr\matl \zeta_0 + \sum_{k=1}^N \tr\matl \zeta_{k|k-1} + \const,
\end{align}
where we used the property that $\det\exp M = \exp\tr M$.
These equations hold for all the parameterizations presented.

The network representation of the complete data \eqref{eq:complete-data_dens} also means its expected log-density factors as
\begin{multline}
  \label{eq:exp_logdens}
  \E_q[\log p_\theta (X_{0:N},y_{0:N})] = \sum_{k=0}^N  E_q[\log p_\theta(y_k|X_{k})] \\
  + E_q[\log p_\theta(X_0)] 
  + \sum_{k=1}^N  E_q[\log p_\theta(X_k|X_{k-1})].
\end{multline}
In general, the expectations do not have closed-form solution, but there exist a plethora of deterministic and stochastic methods to approximate them to any desired order of accuracy, e.g., the unscented transform, Gauss--Hermite quadrature, Monte--Carlo integration, or Quasi-Monte Carlo integration \citep[see, e.g.,][for a comprehensive overview]{sarkka_bayesian_2013}.
All these methods consist of generating samples of a multivariate normal vector and computing a weighted sum of the function, evaluated at each measurement.

To compute the expected log-likelihood of each measurement in \eqref{eq:exp_logdens}, we generate $\nxi$ samples $\{\xi\isamp\}_{i=1}^\nxi$ of an $\nx$-dimensional standard normal vector and their associated weights $\{\omega\isamp\}_{i=1}^\nxi$.
Each sample $x_k\isamp$ of the state and the approximation of the integral are then given by
\begin{align}
  \label{eq:xsamp}
  x_k\isamp &:= \mu_k + S_k\xi\isamp,
  \\
  E_q[\log p_\theta(y_k|X_{k})] &\approx
  \sum_{i=1}^\nxi \omega\isamp\log p_\theta(y_k|x_{k}\isamp).
\end{align}
The same samples and weights can be used to compute the approximation of the expected initial-state log-prior:
\begin{align}
  \label{eq:exp_logprior}
    E_q[\log p_\theta(X_0)] \approx
    \sum_{i=1}^\nxi \omega\isamp\log p_\theta(x_{0}\isamp).
\end{align}

Finally, to compute the expected log-density of each transition in \eqref{eq:exp_logdens}, we generate $\nnu$ samples $\{\nu_{k-1}\isamp, \nu_{k}\isamp\}_{i=1}^\nnu$ of a $2\nx$-dimensional standard normal vector and their associated weights $\{\psi\isamp\}_{i=1}^\nnu$.
The samples  of each state pair are then\footnote{%
  This sampling scheme is the same of \citet{courts_variational_2023}.
}
\begin{subequations}
  \label{eq:sample_pair}
  \begin{align}
    \label{eq:sample_pair_1}
    \tilde x_{k-1}\isamp &:= \mu_{k-1} + S_{k-1}\nu_{k-1}\isamp,
    \\
    \label{eq:sample_pair_2}
    \tilde x_{k}\isamp &:= 
    \mu_{k} + S_{k,k-1}\nu_{k-1}\isamp + S_{k|k-1}\nu_k\isamp,
  \end{align}
\end{subequations}
\noeqref{eq:sample_pair_1,eq:sample_pair_2}%
and the approximation of the expected transition log-density is
\begin{equation}
  \label{eq:exp_trans_logdens}
  E_q[\log p_\theta(X_k|X_{k-1})] \approx
  \sum_{i=1}^\nnu \psi\isamp
  \log p_\theta(\tilde x_{k}\isamp|\tilde x_{k-1}\isamp).
\end{equation}

Substituting the above equations in the ELBO~\eqref{eq:elbo_entro}, with the decomposition of the complete-data density that follows from the state-space representation~\eqref{eq:complete-data_dens} and the parameterizations of the assumed density of Sec.~\ref{sec:param}, we obtain the following approximation for the ELBO, which is used as the objective function:
\begin{multline}
  \label{eq:elbo_approx}
  \elbo(\theta, \cdot) =
  \sum_{i=1}^\nxi \omega\isamp\log p_\theta(x_{0}\isamp)
  + \sum_{k=0}^N\sum_{i=1}^\nxi \omega\isamp\log p_\theta(y_k|x_{k}\isamp)
  \\
  +\sum_{k=1}^N\sum_{i=1}^\nnu \psi\isamp
  \log p_\theta(\tilde x_{k}\isamp|\tilde x_{k-1}\isamp)
  \\
   + \tr\matl \zeta_0 + \sum_{k=1}^N \tr\matl \zeta_{k|k-1} + \const,
\end{multline}
where the function arguments, the decision variables shown in Tab.~\ref{tab:decision}, have been omitted as they depend on the parameterization chosen.
In Eq.~\eqref{eq:elbo_approx}, the samples are computed using \eqref{eq:cov_parameteriz}--\eqref{eq:S_cross_defn}, and \eqref{eq:xsamp}--\eqref{eq:sample_pair}.
Under the steady-state and convolution smoother parameterizations, $\zeta_{k|k-1}=\zcond$ and $S_{k,k-1}=\Scross$ for all $k$.
For the convolution smoother, the mean $\mu$ is computed with \eqref{eq:lin_smoother}.

\section{Optimization}
Two main strategies scale well, in the context of large datasets, for the optimization of the objective function of Sec.~\ref{sec:obj}:
Newton's method with iterative solution of the quadratic subproblem, or stochastic optimization with the dataset divided in mini-batches.
The time-varying parameterization is only practical up to medium-sized datasets, since the storage requirements for the covariances is very high.
The steady-state and convolution smoother parameterizations scale well up to large datasets.
When choosing between both, an important consideration is that the steady-state parameterization yields a sparse Hessian, while the convolution smoother yields a dense.
However, using Hessian-vector products and conjugate-gradient enables Newton's method to be applied to both the sparse and dense parameterizations.
The most straightforward method for computing the objective-function gradient and Hessian-vector products is using automatic differentiation.

For very large datasets, the convolution smoother parameterization is recommended, since the number of decision variables is fixed, independent of the dataset length.
Mini-batching is also recommended in this case, as the abundance of data enables meaningful updates to the decision variables even without processing the full dataset (one epoch).
We assume that the dataset is divided into $\ell$ mini-batches $\{y_{0:N_j}\mbj\}_{j=1}^\ell$ of $N_j+1$ points each.
The mini-batches can come from a single continuous record, split into multiple segments, or from different recording sessions.
In a slight abuse of notation, let $\elbo(\theta, K, \zcond, \Scross; y_{0:N_j}\mbj)$ denote the objective function~\eqref{eq:elbo_approx} with data from the $j$-th mini-batch.
The objective function $\bar\elbo$ of the complete dataset can then be written as
\begin{equation}
  \label{eq:mini-batch-elbo}
  \bar\elbo(\theta, K, \zcond, \Scross) = \sum_{j=1}^\ell
  \elbo(\theta, K, \zcond, \Scross; y_{0:N_j}\mbj),
\end{equation}
and is amenable for optimization by processing each mini-batch in random order, using methods such as stochastic gradient descent or Adam \citep{kingma_adam_2015, schmidt_descending_2021}.
Note that, in this formulation, the interpretation of the initial-state prior is slightly different, as the same prior is used for all mini-batches.
The formulation accepts non-informative or improper priors, however, and the effect of the prior should be neglible for very large datasets.

\section{Simulated Examples}

To illustrate the proposed estimators, we apply them on simulated data of a nonlinear system and large linear systems with large datasets.
All experiments were performed in the WVU Research Computing Thorny Flat HPC cluster and all code used is available as open-source software\footnote{%
  At \url{https://github.com/dimasad/automatica-2024-code}
}.

\subsection{Duffing Oscillator}

The performance of variational system identification with the proposed parameterizations on a nonlinear, non-Gaussian system is demonstrated with the Duffing oscillator, a benchmark model for modeling nonlinear dynamics and chaos \citep{aguirre_modeling_2009}.
The system dynamics is given by the following stochastic differential equation (SDE):
\begin{align}
  \dd X_1(t) &= X_2(t)\,\dd t,\\
  \dd X_2(t) &= [-\alpha X_1(t) -\beta X_1(t) -\delta X_2(t) +\gamma \cos(t)]\,\dd t \\
  &\phantom{===================} + e^\lstdw \,\dd W(t),
\end{align}
where $t\in\reals$ is time, $X_1$ and $X_2$ are the system states, $W$ is a scalar Wiener process, and $\alpha$, $\beta$, $\delta$, $\gamma$ and $\lstdw$ are unknown parameters, to be estimated.
The parameter values in Tab.~\ref{tab:duff-param} were used to  simulate the system with the order \num{1.5} strong Itô--Taylor scheme \citep[Sec.~10.4]{kloeden_numerical_1992} and a time step of $\Tsim$ over the interval $t\in[0,\tfin]$. 
The total simulation length $\tfin$ was varied across experiments to evaluate the estimators as the dataset increases.
Discrete-time, scalar measurements
\begin{equation}
  \label{eq:duff_meas}
  Y_k\sim \normald\big(X_1(k\Ts), e^{2\lstdv}\big),\quad k=0,\dotsc,N
\end{equation}
were drawn with sampling time $\Ts=\num{0.1}$ from a Gaussian distribution with mean given by the simulated state $X_1$ and variance $e^{2\lstdv}$.

\begin{table}
  \centering
  \caption{Parameter values used to generate the data of the Duffing oscillator simulated experiment.}
  \label{tab:duff-param}
  \pgfplotstabletypeset[
    fixed,
    columns/Parameter/.style={column name=Param., string type},
    columns/a/.style={column name=$\alpha$, precision=0},
    columns/b/.style={column name=$\beta$, precision=0},
    columns/d/.style={column name=$\delta$, precision=1},
    columns/g/.style={column name=$\gamma$, precision=1},
    columns/lstdw/.style={column name=$\lstdw$, fixed zerofill, precision=1},
    columns/lstdv/.style={column name=$\lstdv$, fixed zerofill, precision=0},
    columns/Tsim/.style={column name=$\Tsim$, precision=3},
    columns/Ts/.style={column name=$\Ts$, precision=1},
  ]{
    Parameter a  b  d    g    lstdw  lstdv  Tsim  Ts
    Value     1 -1  0.2  0.3  -2.30  -3.00  0.025 0.1
  }
\end{table}

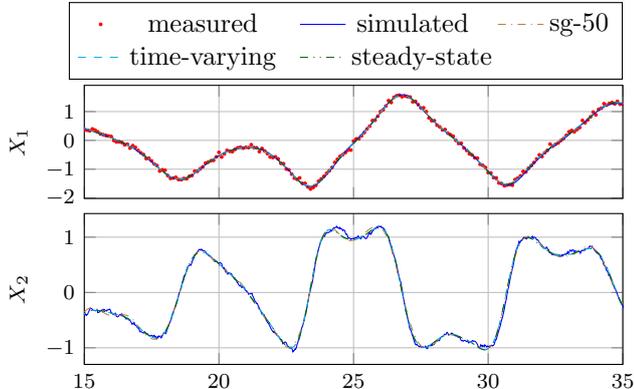
\begin{figure}
  \flushright
  \tikzsetnextfilename{duffing-time-hist.tikz}%
  \tikzpicturedependsonfile{duffing-time-hist.tikz.tex}%
  \begin{tikzpicture}
  \begin{groupplot}[
      group style={
        group size=1 by 2, 
        x descriptions at=edge bottom,
        y descriptions at=edge left,
        horizontal sep=2mm,
        vertical sep=2mm,
      }, 
      height=1.5cm, width=0.8\columnwidth, scale only axis,
      grid=major, xmin=15, xmax=35,
      cycle list={
        {red, mark size=0.5pt, mark=*}, {blue}, {brown, dashdotted},
        {cyan, dashed}, {green!40!black, dashdotdotted},        
      }
    ]
    \pgfplotstableread{duffing-sim.plot}\simtbl
    \pgfplotstableread{duffing-meas.plot}\meastbl
    \pgfplotstableread{duffing-sk-101-xbar.plot}\sktbl
    \pgfplotstableread{duffing-transient-xbar.plot}\transtbl
    \pgfplotstableread{duffing-steady-state-xbar.plot}\sstbl

    \nextgroupplot[
      legend columns=3, ylabel=$X_1$,
      legend style={at={(1, 1.05)},anchor=south east},
    ]

    \addplot+[only marks]
      table [x=t, y=y] {\meastbl};
    \addlegendentry{measured~~~}
    \label{plt:time-hist-meas}

    \addplot+[no marks]
      table [x=t, y=x1] {\simtbl};
    \addlegendentry{simulated~~~}
    \label{plt:time-hist-sim}

    \addplot+[no marks]
      table [x=t, y=x1] {\sktbl};
    \addlegendentry{sg-50}
    \label{plt:time-hist-sk}

    \addplot+[no marks]
      table [x=t, y=x1] {\transtbl};
    \addlegendentry{time-varying~~~}
    \label{plt:time-hist-trans}

    \addplot+[no marks]
      table [x=t, y=x1] {\sstbl};
    \addlegendentry{steady-state}
    \label{plt:time-hist-ss}

    \nextgroupplot[height=2cm, ylabel=$X_2$]
    \addplot+[no marks, draw=none]
      table [x=t, y=x2] {\simtbl}; %Dummy measurement

    \addplot+[no marks]
      table [x=t, y=x2] {\simtbl};

    \addplot+[no marks]
      table [x=t, y=x2] {\sktbl};

    \addplot+[no marks]
      table [x=t, y=x2] {\transtbl};

    \addplot+[no marks]
      table [x=t, y=x2] {\sstbl};
  \end{groupplot}
\end{tikzpicture}%

  \caption{%
    Portion of the time-history of one realization of the Duffing oscillator simulated experiment.
    The simulation ground truth (\ref{plt:time-hist-sim}), noisy measurements (\ref{plt:time-hist-meas}) and mean state estimates under the time-varying (\ref{plt:time-hist-trans}), steady-state (\ref{plt:time-hist-ss}), and convolution smoother (\ref{plt:time-hist-sk}) parameterizations are shown for each of the states.
  }
  \label{fig:duffing-time-hist}
\end{figure}

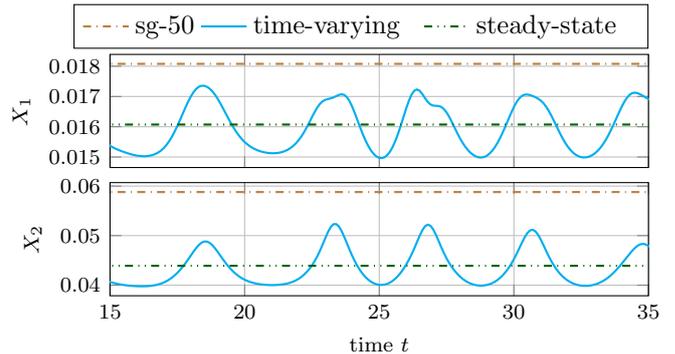
\begin{figure}
  \flushright
  \tikzsetnextfilename{duffing-sigma-hist.tikz}%
  \tikzpicturedependsonfile{duffing-sigma-hist.tikz.tex}%
  \begin{tikzpicture}
  \begin{groupplot}[
      group style={
        group size=1 by 2, 
        x descriptions at=edge bottom,
        y descriptions at=edge left,
        horizontal sep=2mm,
        vertical sep=2mm,
      }, 
      height=1.5cm, width=0.8\columnwidth, scale only axis,
      grid=major, xmin=15, xmax=35,
      cycle list={
        {brown, dashdotted}, {cyan}, {green!40!black, dashdotdotted},
      }
    ]
    \pgfplotstableread{duffing-sim.plot}\simtbl
    \pgfplotstableread{duffing-meas.plot}\meastbl
    \pgfplotstableread{duffing-sk-101-xbar.plot}\sktbl
    \pgfplotstableread{duffing-transient-xbar.plot}\transtbl
    \pgfplotstableread{duffing-steady-state-xbar.plot}\sstbl
    
    \nextgroupplot[
      ylabel=$X_1$, 
      legend columns=5, legend style={at={(1, 1.05)},anchor=south east},
      tick label style={
        /pgf/number format/fixed, /pgf/number format/precision=3
      },
    ]
    \addplot+[no marks, thick]
      table [x=t, y=stdx1] {\sktbl};
    \addlegendentry{sg-50}

    \addplot+[no marks, thick]
      table [x=t, y=stdx1] {\transtbl};
    \addlegendentry{time-varying~~~}

    \addplot+[no marks, thick]
      table [x=t, y=stdx1] {\sstbl};
    \addlegendentry{steady-state~~~}

    \nextgroupplot[
      ylabel=$X_2$, xlabel=time $t$,
      tick label style={/pgf/number format/fixed}
    ]
    \addplot+[no marks, thick]
      table [x=t, y=stdx2] {\sktbl};
    \addplot+[no marks, thick]
      table [x=t, y=stdx2] {\transtbl};
    \addplot+[no marks, thick]
      table [x=t, y=stdx2] {\sstbl};

  \end{groupplot}
\end{tikzpicture}%

  \caption{%
    Standard deviation of the estimated state posterior for the realization of the Duffing oscillator simulated experiment shown in Fig.~\ref{fig:duffing-time-hist}.
  }
  \label{fig:duffing-sigma-hist}
\end{figure}

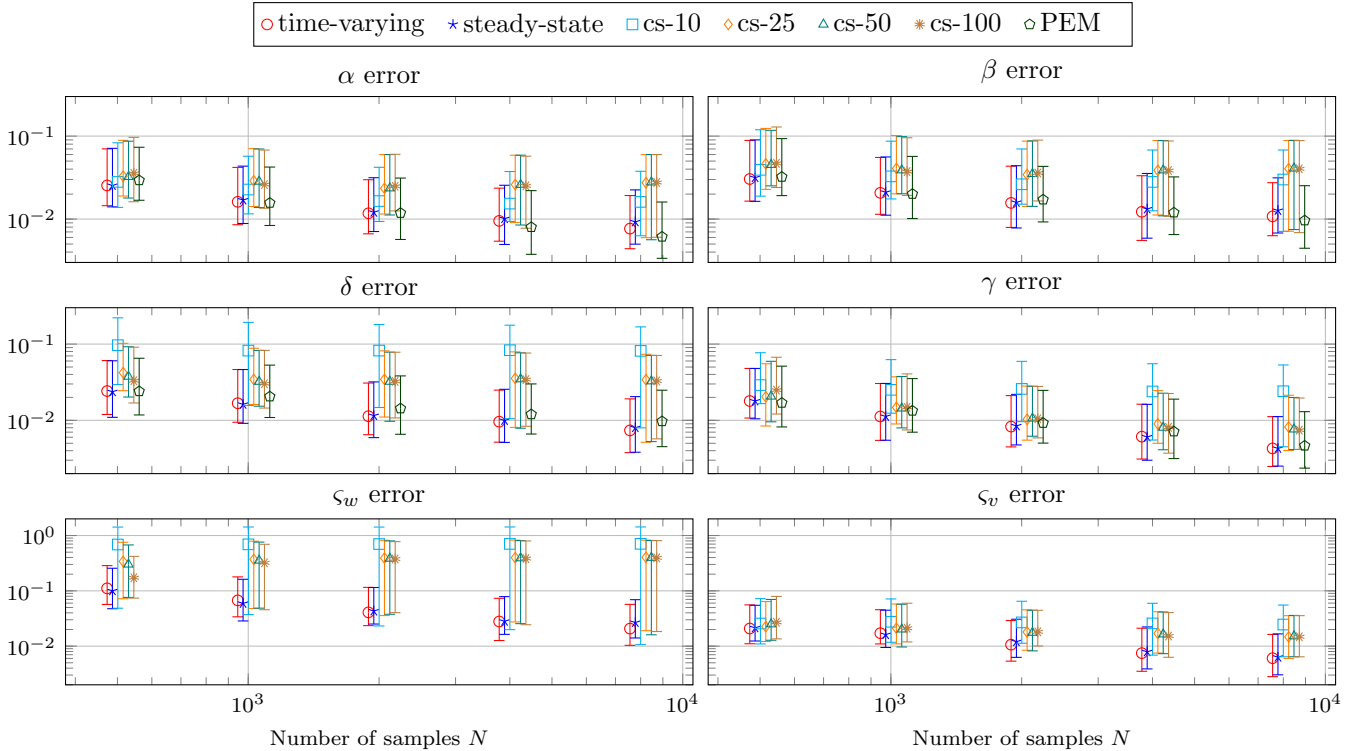
\begin{figure*}
  \centering
  \tikzsetnextfilename{duffing-perr.tikz}%
  \tikzpicturedependsonfile{duffing-perr.tikz.tex}%
  \begin{tikzpicture}
  \begin{groupplot}[
      group style={
        group size=2 by 3, 
        x descriptions at=edge bottom,
        y descriptions at=edge left,
        horizontal sep=2mm,
        vertical sep=0.6cm,
      }, 
      height=2.2cm, width=0.45\textwidth, scale only axis,
      xmode=log, ymode=log, grid=major, ymax=0.3, ymin=0.002,
      error bars/y dir=both, error bars/y explicit,
      cycle list={
        {red, mark=o}, {blue, mark=star}, {cyan, mark=square}, 
        {orange, mark=diamond}, {teal, mark=triangle}, 
        {brown, mark=10-pointed star}, {green!25!black, mark=pentagon},
      }
    ]
    \pgfplotstableread{duffing-transient.plot}\transtbl
    \pgfplotstableread{duffing-steady-state.plot}\sstbl
    \pgfplotstableread{duffing-sk-21.plot}\skatbl
    \pgfplotstableread{duffing-sk-51.plot}\skbtbl
    \pgfplotstableread{duffing-sk-101.plot}\skctbl
    \pgfplotstableread{duffing-sk-201.plot}\skdtbl
    \pgfplotstableread{duffing-PEM.plot}\pemtbl

    \nextgroupplot[
      legend columns=7, 
      legend style={at={(1, 1.3)},anchor=south},
      title=$\alpha$ error, ymin=0.003
    ]
    \addplot+[only marks, xshift=-4pt] 
      table [x=N_med, y=a_med, y error minus=a_low, y error plus=a_high] 
      {\transtbl};
    \addlegendentry{time-varying~~~}

    \addplot+[only marks, xshift=-2pt] 
      table [x=N_med, y=a_med, y error minus=a_low, y error plus=a_high] 
      {\sstbl};
    \addlegendentry{steady-state~~~}

    \addplot+[only marks, xshift=0pt] 
      table [x=N_med, y=a_med, y error minus=a_low, y error plus=a_high] 
      {\skatbl};
    \addlegendentry{cs-10~~~}

    \addplot+[only marks, xshift=2pt] 
      table [x=N_med, y=a_med, y error minus=a_low, y error plus=a_high] 
      {\skbtbl};
    \addlegendentry{cs-25~~~}

    \addplot+[only marks, xshift=4pt] 
      table [x=N_med, y=a_med, y error minus=a_low, y error plus=a_high] 
      {\skctbl};
    \addlegendentry{cs-50~~~}

    \addplot+[only marks, xshift=6pt] 
      table [x=N_med, y=a_med, y error minus=a_low, y error plus=a_high] 
      {\skdtbl};
    \addlegendentry{cs-100~~~}

    \addplot+[only marks, xshift=8pt] 
      table [x=N_med, y=a_med, y error minus=a_low, y error plus=a_high] 
      {\pemtbl};
    \addlegendentry{PEM~~~}

    \nextgroupplot[title=$\beta$ error, ymin=0.003]
    \addplot+[only marks, xshift=-4pt] 
      table [x=N_med, y=b_med, y error minus=b_low, y error plus=b_high] 
      {\transtbl};

    \addplot+[only marks, xshift=-2pt] 
      table [x=N_med, y=b_med, y error minus=b_low, y error plus=b_high] 
      {\sstbl};

    \addplot+[only marks, xshift=0pt] 
      table [x=N_med, y=b_med, y error minus=b_low, y error plus=b_high] 
      {\skatbl};

    \addplot+[only marks, xshift=2pt] 
      table [x=N_med, y=b_med, y error minus=b_low, y error plus=b_high] 
      {\skbtbl};

    \addplot+[only marks, xshift=4pt] 
      table [x=N_med, y=b_med, y error minus=b_low, y error plus=b_high] 
      {\skctbl};

    \addplot+[only marks, xshift=6pt] 
      table [x=N_med, y=b_med, y error minus=b_low, y error plus=b_high] 
      {\skdtbl};

    \addplot+[only marks, xshift=8pt] 
      table [x=N_med, y=b_med, y error minus=b_low, y error plus=b_high] 
      {\pemtbl};

    \nextgroupplot[title=$\delta$ error]
    \addplot+[only marks, xshift=-4pt] 
      table [x=N_med, y=d_med, y error minus=d_low, y error plus=d_high] 
      {\transtbl};

    \addplot+[only marks, xshift=-2pt] 
      table [x=N_med, y=d_med, y error minus=d_low, y error plus=d_high] 
      {\sstbl};

    \addplot+[only marks, xshift=0pt] 
      table [x=N_med, y=d_med, y error minus=d_low, y error plus=d_high] 
      {\skatbl};

    \addplot+[only marks, xshift=2pt] 
      table [x=N_med, y=d_med, y error minus=d_low, y error plus=d_high] 
      {\skbtbl};

    \addplot+[only marks, xshift=4pt] 
      table [x=N_med, y=d_med, y error minus=d_low, y error plus=d_high] 
      {\skctbl};

    \addplot+[only marks, xshift=6pt] 
      table [x=N_med, y=d_med, y error minus=d_low, y error plus=d_high] 
      {\skdtbl};

    \addplot+[only marks, xshift=8pt] 
      table [x=N_med, y=d_med, y error minus=d_low, y error plus=d_high] 
      {\pemtbl};

    \nextgroupplot[title=$\gamma$ error]
    \addplot+[only marks, xshift=-4pt] 
      table [x=N_med, y=g_med, y error minus=g_low, y error plus=g_high] 
      {\transtbl};

    \addplot+[only marks, xshift=-2pt] 
      table [x=N_med, y=g_med, y error minus=g_low, y error plus=g_high] 
      {\sstbl};

    \addplot+[only marks, xshift=0pt] 
      table [x=N_med, y=g_med, y error minus=g_low, y error plus=g_high] 
      {\skatbl};

    \addplot+[only marks, xshift=2pt] 
      table [x=N_med, y=g_med, y error minus=g_low, y error plus=g_high] 
      {\skbtbl};

    \addplot+[only marks, xshift=4pt] 
      table [x=N_med, y=g_med, y error minus=g_low, y error plus=g_high] 
      {\skctbl};

    \addplot+[only marks, xshift=6pt] 
      table [x=N_med, y=g_med, y error minus=g_low, y error plus=g_high] 
      {\skdtbl};

    \addplot+[only marks, xshift=8pt] 
      table [x=N_med, y=g_med, y error minus=g_low, y error plus=g_high] 
      {\pemtbl};

    \nextgroupplot[
      title=$\varsigma_w$ error, ymax=2, xlabel=Number of samples $N$,
    ]
    \addplot+[only marks, xshift=-4pt] 
      table [x=N_med, y=logstd_w_med, y error minus=logstd_w_low, y error plus=logstd_w_high] 
      {\transtbl};

    \addplot+[only marks, xshift=-2pt] 
      table [x=N_med, y=logstd_w_med, y error minus=logstd_w_low, y error plus=logstd_w_high] 
      {\sstbl};

    \addplot+[only marks, xshift=0pt] 
      table [x=N_med, y=logstd_w_med, y error minus=logstd_w_low, y error plus=logstd_w_high] 
      {\skatbl};

    \addplot+[only marks, xshift=2pt] 
      table [x=N_med, y=logstd_w_med, y error minus=logstd_w_low, y error plus=logstd_w_high] 
      {\skbtbl};

    \addplot+[only marks, xshift=4pt] 
      table [x=N_med, y=logstd_w_med, y error minus=logstd_w_low, y error plus=logstd_w_high] 
      {\skctbl};

    \addplot+[only marks, xshift=6pt] 
      table [x=N_med, y=logstd_w_med, y error minus=logstd_w_low, y error plus=logstd_w_high] 
      {\skdtbl};

    \nextgroupplot[
      title=$\varsigma_v$ error, ymax=2, xlabel=Number of samples $N$,
    ]
    \addplot+[only marks, xshift=-4pt] 
      table [x=N_med, y=logstd_v_med, y error minus=logstd_v_low, y error plus=logstd_v_high] 
      {\transtbl};

    \addplot+[only marks, xshift=-2pt] 
      table [x=N_med, y=logstd_v_med, y error minus=logstd_v_low, y error plus=logstd_v_high] 
      {\sstbl};

    \addplot+[only marks, xshift=0pt] 
      table [x=N_med, y=logstd_v_med, y error minus=logstd_v_low, y error plus=logstd_v_high] 
      {\skatbl};

    \addplot+[only marks, xshift=2pt] 
      table [x=N_med, y=logstd_v_med, y error minus=logstd_v_low, y error plus=logstd_v_high] 
      {\skbtbl};

    \addplot+[only marks, xshift=4pt] 
      table [x=N_med, y=logstd_v_med, y error minus=logstd_v_low, y error plus=logstd_v_high] 
      {\skctbl};

    \addplot+[only marks, xshift=6pt] 
      table [x=N_med, y=logstd_v_med, y error minus=logstd_v_low, y error plus=logstd_v_high] 
      {\skdtbl};
  \end{groupplot}
\end{tikzpicture}%

  \caption{%
    Estimation error of each parameter of the the Duffing oscillator experiment for different dataset sizes.
    The error bars indicate the first and third quartiles while the marker indicates the median of equal simulations with different random seeds.
    All methods were evaluated at the same number of samples $N\in\{500, 1000, 2000, 4000, 8000\}$ but the markers staggered horizontally, slightly, to improve legibility.
    The cs-$M$ entries correspond to the convolution smoother parameterization with for a window length of $M$ samples.
  }
  \label{fig:duffing-perr}
\end{figure*}
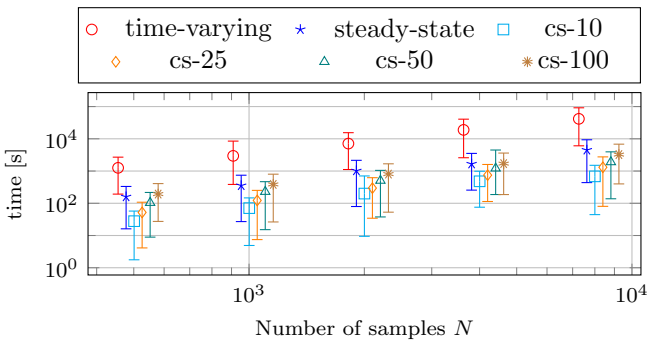
\begin{figure}
  \centering
  \tikzsetnextfilename{duffing-t.tikz}%
  \tikzpicturedependsonfile{duffing-t.tikz.tex}%
  \begin{tikzpicture}
  \begin{loglogaxis}[
      height=4cm, width=\columnwidth,
      xmode=log, ymode=log, grid=major, 
      minor ytick={1e1,1e3,1e5}, yminorgrids=true,
      xlabel=Number of samples $N$, ylabel={time [s]},
      error bars/y dir=both, error bars/y explicit,
      legend columns=3,
      legend style={at={(1, 1.05)},anchor=south east},
      cycle list={
        {red, mark=o}, {blue, mark=star}, {cyan, mark=square}, 
        {orange, mark=diamond}, {teal, mark=triangle}, 
        {brown, mark=10-pointed star}
      }
    ]
    \pgfplotstableread{duffing-transient.plot}\transtbl
    \pgfplotstableread{duffing-steady-state.plot}\sstbl
    \pgfplotstableread{duffing-sk-21.plot}\skatbl
    \pgfplotstableread{duffing-sk-51.plot}\skbtbl
    \pgfplotstableread{duffing-sk-101.plot}\skctbl
    \pgfplotstableread{duffing-sk-201.plot}\skdtbl
    
    \addplot+[only marks, xshift=-6pt] 
      table [x=N_med, y=opt_time_med, y error minus=opt_time_low, y error plus=opt_time_high] 
      {\transtbl};
    \addlegendentry{time-varying~~~}

    \addplot+[only marks, xshift=-3pt] 
      table [x=N_med, y=opt_time_med, y error minus=opt_time_low, y error plus=opt_time_high] 
      {\sstbl};
    \addlegendentry{steady-state~~~}

    \addplot+[only marks, xshift=0pt] 
      table [x=N_med, y=opt_time_med, y error minus=opt_time_low, y error plus=opt_time_high] 
      {\skatbl};
    \addlegendentry{cs-10~~~}

    \addplot+[only marks, xshift=3pt] 
      table [x=N_med, y=opt_time_med, y error minus=opt_time_low, y error plus=opt_time_high] 
      {\skbtbl};
    \addlegendentry{cs-25~~~}

    \addplot+[only marks, xshift=6pt] 
      table [x=N_med, y=opt_time_med, y error minus=opt_time_low, y error plus=opt_time_high] 
      {\skctbl};
    \addlegendentry{cs-50~~~}

    \addplot+[only marks, xshift=9pt] 
      table [x=N_med, y=opt_time_med, y error minus=opt_time_low, y error plus=opt_time_high] 
      {\skdtbl};
    \addlegendentry{cs-100~~~}
  \end{loglogaxis}
\end{tikzpicture}%

  \caption{%
    Time needed to compute the estimate of the duffing experiment for different dataset sizes.
    The labels and horizontal stagger are the same as in Fig.~\ref{fig:duffing-perr}.
  }
  \label{fig:duffing-t}
\end{figure}

A total of~200 independent simulations of the system were made for each of the experiment time lengths of $\tfin\in\{50, 100, 200, 400, 800\}$ and the proposed estimators were applied to the data.
The model used for estimation was discretized with the same SDE scheme used for simulation, but with only one time-step between measurements.
Second-order Gauss--Hermite nodes and weights were used for quadrature in \eqref{eq:xsamp} and~\eqref{eq:exp_trans_logdens}.
The convolution smoother parameterization was applied with the window lengths of $M\in\{10, 25, 50, 100\}$.
Noninformative improper uniform priors were used for all parameters and initial states.
In addition, all decision variables were initialized at zero.
The \texttt{trust-constr} method of \texttt{scipy.optimize.minimize} \citep{virtanen_scipy_2020} was used to solve the underlying optimization problem with the gradient and Hessian--vector product computed using automatic differentiation.

The proposed estimators were compared with the prediction error method (PEM) using a steady-state extended Kalman filter (EKF) in innovation form \citep[Chap.~5]{maine_formulation_1981, jategaonkar_flight_2015}.
As this method is not robust to null initialization, the parameters were initialized at their true values.
The same optimization method as for GVI was used for solving the underlying maximization problem.

With the parameters chosen, the system has a double-well potential which, with the forcing function used, exhibits dynamical chaos, as shown in Fig.~\ref{fig:duffing-time-hist}.
This means that the state is not restricted to small deviations from an equilibrium condition and that the smoothed state posterior covariance does not converge to a single point, oscillating instead.
This behavior can be observed in Fig.~\ref{fig:duffing-sigma-hist}, which shows a portion of the estimated state standard deviation obtained with different parameterizations.
The experiments show that while the time-varying parameterization captures the state covariance better than the steady-state one, the error of their parameter estimates is comparable, as evidenced in Fig.~\ref{fig:duffing-perr}.
Due to the reduced number of parameters and its simplified objective function, however, the steady-state estimates can be computed an order of magnitude faster than when the time-varying estimator is used, as shown in Fig.~\ref{fig:duffing-t}.

While PEM attained nearly identical performance to the steady-state and time-varying parameterizations of GVI, it was initialized at the true parameters of the associated steady-state EKF.
As the system is chaotic, small errors in the predictor can cause the predictor to diverge from the data, causing the optimization to stop.
All of the proposed parameterizations of GVI were able to converge from null--random initial guesses to the true parameters, however.
This shows the method is a promising alternative to PEM for estimating chaotic and unstable systems without requiring an initial guess.

\subsection{Linear Systems with Large Datasets}

\begin{figure*}
  \centering
  \tikzsetnextfilename{linsys-eratio.tikz}%
  \tikzpicturedependsonfile{linsys-eratio.tikz.tex}%
  \begin{tikzpicture}
  \begin{groupplot}[
      group style={
        group size=1 by 1, 
        vertical sep=2.3cm,
      },
      boxplot, boxplot/draw direction=y,
      ymode=log, ytick={1e-4, 1e-3, 1e-2, 1e-1, 1e0, 1e1},
      height=4cm, width=0.95\textwidth,
      ylabel={IER},
      grid=major,
      xtick={1,...,8}, xticklabel style={align=center},
      cycle list={red,blue,black,brown,teal,orange,violet,cyan},
    ]
    
    \nextgroupplot[
      title={Linear system impulse response error ratio with mini-batching},
      xticklabels={
        GVI\\$\nx=5$\\$\ninp=2$\\$\ny=2$,    PEM\\$\nx=5$\\$\ninp=2$\\$\ny=2$,
        GVI\\$\nx=10$\\$\ninp=4$\\$\ny=4$,   PEM\\$\nx=10$\\$\ninp=4$\\$\ny=4$,
        GVI\\$\nx=20$\\$\ninp=8$\\$\ny=8$,   PEM\\$\nx=20$\\$\ninp=8$\\$\ny=8$,
        GVI\\$\nx=40$\\$\ninp=16$\\$\ny=16$, PEM\\$\nx=40$\\$\ninp=16$\\$\ny=16$,
      },
      ymin=1e-4, table/y index=5, 
      mark size=1pt,
    ]
    \addplot+ table[] {linsys-1-gvi.plot};
    \addplot+ table[] {linsys-1-pem.plot};
    \addplot+ table[] {linsys-2-gvi.plot};
    \addplot+ table[] {linsys-2-pem.plot};
    \addplot+ table[] {linsys-4-gvi.plot};
    \addplot+ table[] {linsys-4-pem.plot};
    \addplot+ table[] {linsys-8-gvi.plot};
    \addplot+ table[] {linsys-8-pem.plot};
  \end{groupplot}
\end{tikzpicture}%

  \caption{%
    Box-and-whisker plots of the impulse error ratio (IER) of the linear system simulation experiment, as defined in \eqref{eq:ier}, using the Adam optimizer and the full dataset split into \num{1000} mini-batches of \num{10000} time samples each.
  }
  \label{fig:linsys-eratio}
\end{figure*}
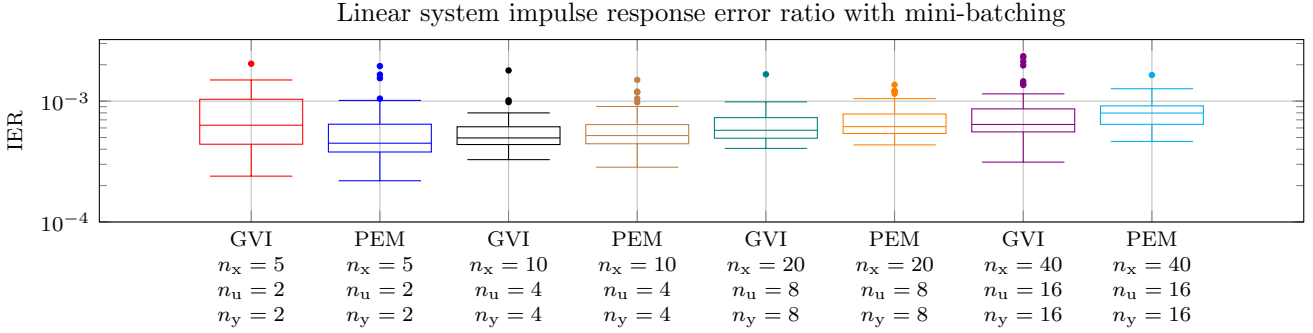

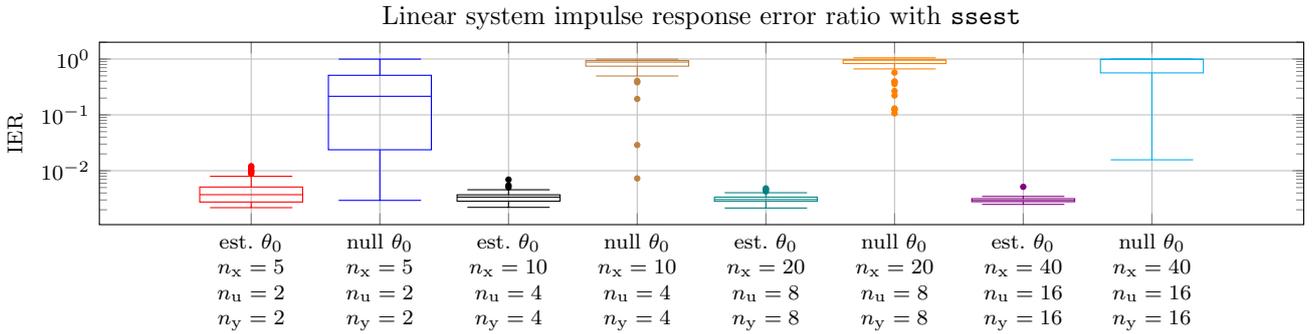
\begin{figure*}
  \centering
  \tikzsetnextfilename{ssest-eratio.tikz}%
  \tikzpicturedependsonfile{ssest-eratio.tikz.tex}%
  \begin{tikzpicture}
  \begin{groupplot}[
      group style={
        group size=1 by 1, 
        vertical sep=2.3cm,
      },
      boxplot, boxplot/draw direction=y,
      ymode=log, ytick={1e-4, 1e-3, 1e-2, 1e-1, 1e0, 1e1},
      height=4cm, width=0.95\textwidth,
      ylabel={IER},
      grid=major,
      xtick={1,...,8}, xticklabel style={align=center},
      cycle list={red,blue,black,brown,teal,orange,violet,cyan},
    ]
    
    \nextgroupplot[
      title={Linear system impulse response error ratio with \texttt{ssest}},
      xticklabels={
        est.\ $\theta_0$\\$\nx=5$\\$\ninp=2$\\$\ny=2$,    null $\theta_0$\\$\nx=5$\\$\ninp=2$\\$\ny=2$,
        est.\ $\theta_0$\\$\nx=10$\\$\ninp=4$\\$\ny=4$,   null $\theta_0$\\$\nx=10$\\$\ninp=4$\\$\ny=4$,
        est.\ $\theta_0$\\$\nx=20$\\$\ninp=8$\\$\ny=8$,   null $\theta_0$\\$\nx=20$\\$\ninp=8$\\$\ny=8$,
        est.\ $\theta_0$\\$\nx=40$\\$\ninp=16$\\$\ny=16$, null $\theta_0$\\$\nx=40$\\$\ninp=16$\\$\ny=16$,
      },
      ymax=2, table/y index=2, 
      mark size=1pt,
    ]
    \addplot+ table[] {linsys-1-initialized.plot};
    \addplot+ table[] {linsys-1-nullsys0.plot};
    \addplot+ table[] {linsys-2-initialized.plot};
    \addplot+ table[] {linsys-2-nullsys0.plot};
    \addplot+ table[] {linsys-4-initialized.plot};
    \addplot+ table[] {linsys-4-nullsys0.plot};
    \addplot+ table[] {linsys-8-initialized.plot};
    \addplot+ table[] {linsys-8-nullsys0.plot};
  \end{groupplot}
\end{tikzpicture}%

  \caption{%
    Box-and-whisker plots of the impulse error ratio (IER) of the linear system simulation experiment, as defined in \eqref{eq:ier}, using MATLAB's \texttt{ssest}, with only a single batch of \num{10000} time samples.
    The label ``null $\theta_0$'' indicates that the model was initialized to null--random values and ``est. $\theta_0$'' that an initial guess was computed using an automatically-chosen auxiliary estimator.
  }
  \label{fig:ssest-eratio}
\end{figure*}

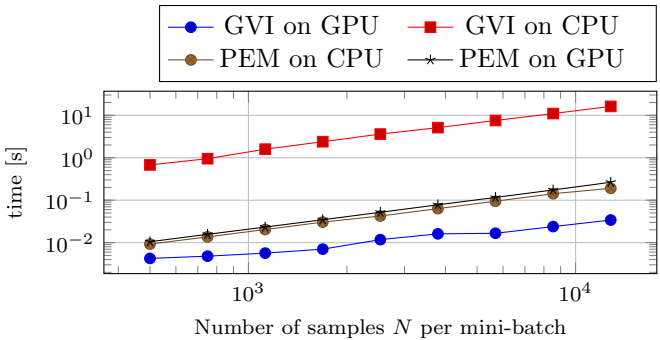
\begin{figure}
  \centering
  \tikzsetnextfilename{linsys-t.tikz}%
  \tikzpicturedependsonfile{linsys-t.tikz.tex}%
  \begin{tikzpicture}
  \begin{loglogaxis}[
      height=4cm, width=\columnwidth,
      xmode=log, ymode=log, grid=major, 
      xlabel=Number of samples $N$, ylabel={time [s]},
      ytick={1e-3, 1e-2, 1e-1, 1e0, 1e1, 1e2},
      legend columns=2, legend style={at={(1, 1.05)},anchor=south east},
      %cycle list={
      %  {red, mark=o}, {blue, mark=star}, {cyan, mark=square}, 
      %  {orange, mark=diamond}, {teal, mark=triangle}, 
      %  {brown, mark=10-pointed star}
      %}
    ]
    %\pgfplotstableread{linsys-time-0-pem-cpu.plot}\smallpemcputbl
    %\pgfplotstableread{linsys-time-0-pem-gpu.plot}\smallpemgputbl
    %\pgfplotstableread{linsys-time-0-gvi-cpu.plot}\smallgvicputbl
    %\pgfplotstableread{linsys-time-0-gvi-gpu.plot}\smallgvigputbl
    \pgfplotstableread{linsys-time-1-pem-cpu.plot}\largepemcputbl
    \pgfplotstableread{linsys-time-1-pem-gpu.plot}\largepemgputbl
    \pgfplotstableread{linsys-time-1-gvi-cpu.plot}\largegvicputbl
    \pgfplotstableread{linsys-time-1-gvi-gpu.plot}\largegvigputbl
    
    \addplot+ table [x index=3, y index=5] {\largegvigputbl};
    \addlegendentry{GVI on GPU~~~}

    \addplot+ table [x index=3, y index=5] {\largegvicputbl};
    \addlegendentry{GVI on CPU~~~}

    \addplot+ table [x index=3, y index=5] {\largepemcputbl};
    \addlegendentry{PEM on CPU~~~}

    \addplot+ table [x index=3, y index=5] {\largepemgputbl};
    \addlegendentry{PEM on GPU~~~}
  \end{loglogaxis}
\end{tikzpicture}%

  \caption{%
    Time need to run compute the objective function and its gradient, for the linear system example, with different numbers of samples $N$.
  }
  \label{fig:linsys-t}
\end{figure}

For linear--Gaussian systems, variational system identification and prediction-error methods are different ways to represent nearly identical estimation problems.
The differences in the decision variables and objective function of their optimization problems, however, gives both of these problems different convergence properties and computational complexity.
Important properties of the proposed convolution smoother formulation of variational system identification for large datasets are that it is less susceptible to local minima, its objective function is embarrassingly parallelizable, and can exploit the computational power of modern GPUs.
These properties are demonstrated with simulated examples in this section.

For each realization of the experiment, random $A$, $B$, $C$, $D$ matrices of the appropriate sizes were drawn, with all eigenvalues $\lambda_i$ of $A$ satisfying $|\lambda_i|<\num{0.95}$.
The linear discrete-time system was then simulated with zero initial state $X_0$, a random binary signal as the known input $u_k$, and independent multivariate Gaussian white noise vectors $W_k$ and $V_k$, according to
\begin{align}
  X_{k+1} &= A X_k + Bu_k + W_k, \\ 
  Y_{k} &= C X_k + Du_k + V_k.
\end{align}
For evaluating system identification of large systems from large datasets, each realization of the experiment consisted of $N=10^7$ samples, with the number of states varying from $n_x=5$ to $n_x=40$.

Variational system identification was then applied to the data for estimating a linear--Gaussian system using the convolution smoother formulation.
The window length of $M=50$ was used for all experiments, which captures well the maximum pole radius of the $A$ matrices that were drawn.
The data was divided into \num{1000} mini-batches of \num{10000} points each and the Adam optimizer \citep{kingma_adam_2015} was used to minimize the ELBO, starting from the initial guess of zero for all parameters except the central smoother gain $K^{(0)}$, which was initialized near zero with independent Gaussian random values with variance \num{1e-6}.
An exponential learning rate decay was used and the optimizer was run for a total of 75 epochs.
We note that, since all parameters of the system matrices were kept free, the system is overparameterized \citep{mckelvey_data_2004} but this is not a problem for the stochastic optimization with Adam.

To evaluate the estimates, the impulse response of the system with the true and estimated parameters was compared over a horizon of $N_h=100$ time samples.
Letting $\imptrue$ denote the response of true system output $i$ to a discrete-time unit impulse in input $j$, at time sample $k$, and $\impest$ the same response for the estimated system, we define the $\ell^2$-norm impulse error ratio (IER) as
\begin{equation}
  \label{eq:ier}
  \operatorname{IER} = 
  \frac{1}{\ny\ninp}\sum_{i=1}^\ny\sum_{j=1}^\ninp
  \frac{
    \sqrt{\sum_{k=0}^{N_h}\left(\imptrue-\impest\right)^2}
  }{
    \sqrt{\sum_{k=0}^{N_h}\left(\imptrue\right)^2}
  },
\end{equation}
that is, the average, over all input-output pairs, of the ratio between the $\ell^2$-norm of the impulse response error and the $\ell^2$-norm of the true system impulse response signal.

The PEM using a steady-state linear state-space predictor in innovations form \citep{mckelvey_data_2004, wills_gradient-based_2008} was used to benchmark the results, using three different approaches.
To compare with the established approach, MATLAB's \texttt{ssest} function, from the System Identification Toolbox, was used with a single batch of \num{10000} time samples.
The estimator was called with two options: estimating the initial system using an automatically chosen auxiliary method and with null $A$, $B$, $C$, $D$ matrices and random near-zero $K$ gain matrices.
A final approach for PEM consisted of using the stochastic Adam optimizer to minimize the prediction errors of the full dataset, split into mini-batches like GVI, starting from null system matrices and random near-zero gain matrices.

The results of a total of 50 independent simulations for each of the system sizes evaluated are shown in Figs.~\ref{fig:linsys-eratio} and~\ref{fig:ssest-eratio}.
The IER metric of Fig.~\ref{fig:linsys-eratio} shows that the estimated systems converged to the true system impulse respose, with an error ratio below \SI{0.3}{\%} for all cases, even with null--random initial guesses.
Interestingly, the use of stochastic optimization and mini-batching can also enable the PEM to converge to the true solution from null--random initial guesses as well, something that does not occur when using \texttt{ssest}, as shown in Fig.~\ref{fig:ssest-eratio}.
A notable limitation of PEM is that it some parameters can result in unstable predictors, which leads to premature stopping of the optimization and may require additional tuning.
This issue does not arise in GVI, however, since convolution smoothers are inherently bounded-input bounded-output stable.

Computationally, GVI is embarrassingly parallel and PEM is sequential.
This means that, even though more mathematical operations are performed in GVI then PEM, as evidenced by the time in Fig.~\ref{fig:linsys-t} it takes to run both in CPUs, GVI runs two orders of magnitude faster on a GPU than on a CPU.
PEM, on the other hand, ran slightly slower on the GPU than on CPU.

A major limitation of the traditional approach, examplified by \texttt{ssest}, to applying PEM to estimate large models from large datasets is that initial guesses are necessary but subspace methods for obtaining them do not scale well.
To illustrate that, GVI was given datasets with \num{1000} times more time samples than \texttt{ssest}, but was able to compute each solution in less than 1 hour and 15 minutes.
For the experiments with $\nx=40$ states, $\ninp=16$ inputs, and $\ny=16$ outputs, \texttt{ssest} took over 4 hours and 38 minutes to compute each estimate.
The use of more data enabled lower errors to be achieved by GVI and PEM, as seen by comparing Figs.~\ref{fig:linsys-eratio} and~\ref{fig:ssest-eratio}.

\section{Conclusion and Future Work}

This paper shows how recognizing the Bayesian network topology of the state-path posterior distribution can enable formulating the variational system identification problem of \citet{courts_variational_2021, courts_variational_2023} as an unconstrained optimization.
The network representation also leads, naturally, to steady-state and convolution smoother representations that enable the identification of large models from large datasets with null and random initial guesses, without convergence to local optima.
The proposed estimators outperform the state of the art for both nonlinear and linear system identification.

The results are obtained by bringing together ideas from machine learning and systems and control.
Applications in machine learning have driven much of the recent developments in variational inference, with variational autoencoders becoming a major architecture for deep learning \citep{zhang_advances_2019}.
Stochastic optimization with mini-batching is the only tractable approach for estimating parameters of deep models from large datasets, and Adam has become one of the most established methods \citep{schmidt_descending_2021}.
A notable property of these methods is that they can converge from random initial guesses to a set of near-equivalent solutions, as the models are massively overparameterized.
Parallelism and architectures which perform well on GPUs are favored due to speedups in training.

The estimators introduced in this paper are embarassingly parallel, can use mini-batching, are amenable to stochastic optimization, perform well on GPUs, can be applied to overparameterized models, and converge to near-equivalent near-optimal solutions when started at null--random guesses.
These were only possible by applying techniques of square-root Kalman filtering \citep[Chap.~12]{kailath_linear_2000} and theoretical properties of state-space Bayesian networks.
Beyond the applications presented in this paper, the proposed methods are well-suited for identifying deep models for applications in systems and control \citep{gedon_deep_2021}.
In addition, noting that training of recurrent neural networks can be posed as system identification \citep{ribeiro_beyond_2020, ribeiro_smoothness_2020}, they might be of value for learning chaotic recurrent networks in a stable, parallel manner.

A shortcoming of the proposed convolution-smoother parameterization is its limited expressive power, however, which results in larger errors than PEM and the time-varying GVI.
For non-Gaussian, nonlinear system identification a promising avenue is using a convolution neural network as the smoother $\phi$ in \eqref{eq:smoother_fcn}.
This has the potential to enable the estimation of large nonlinear models from large datasets without incurring in much performance loss with respect to the time-varying parameterization.

%\nolinenumbers
\bibliography{bib}

\begin{thebibliography}{46}
\expandafter\ifx\csname natexlab\endcsname\relax\def\natexlab#1{#1}\fi
\providecommand{\url}[1]{\texttt{#1}}
\providecommand{\href}[2]{#2}
\providecommand{\path}[1]{#1}
\providecommand{\DOIprefix}{doi:}
\providecommand{\ArXivprefix}{arXiv:}
\providecommand{\URLprefix}{URL: }
\providecommand{\Pubmedprefix}{pmid:}
\providecommand{\doi}[1]{\href{http://dx.doi.org/#1}{\path{#1}}}
\providecommand{\Pubmed}[1]{\href{pmid:#1}{\path{#1}}}
\providecommand{\bibinfo}[2]{#2}
\ifx\xfnm\relax \def\xfnm[#1]{\unskip,\space#1}\fi
%Type = Article
\bibitem[{Aguirre \& Letellier(2009)}]{aguirre_modeling_2009}
\bibinfo{author}{Aguirre, L.}, \& \bibinfo{author}{Letellier, C.}
  (\bibinfo{year}{2009}).
\newblock \bibinfo{title}{Modeling nonlinear dynamics and chaos: a review}.
\newblock {\it \bibinfo{journal}{Mathematical Problems in Engineering}\/},
  {\it \bibinfo{volume}{2009}\/}, \bibinfo{pages}{238960}.
  \DOIprefix\doi{10.1155/2009/238960}.
%Type = Article
\bibitem[{Al-Mohy \& Higham(2010)}]{al-mohy_new_2010}
\bibinfo{author}{Al-Mohy, A.~H.}, \& \bibinfo{author}{Higham, N.~J.}
  (\bibinfo{year}{2010}).
\newblock \bibinfo{title}{A {New} {Scaling} and {Squaring} {Algorithm} for the
  {Matrix} {Exponential}}.
\newblock {\it \bibinfo{journal}{SIAM Journal on Matrix Analysis and
  Applications}\/},  {\it \bibinfo{volume}{31}\/}, \bibinfo{pages}{970--989}.
  \DOIprefix\doi{10.1137/09074721X}.
%Type = Article
\bibitem[{Andrieu et~al.(2004)Andrieu, Doucet, Singh \&
  Tadić}]{andrieu_particle_2004}
\bibinfo{author}{Andrieu, C.}, \bibinfo{author}{Doucet, A.},
  \bibinfo{author}{Singh, S.}, \& \bibinfo{author}{Tadić, V.}
  (\bibinfo{year}{2004}).
\newblock \bibinfo{title}{Particle methods for change detection, system
  identification, and control}.
\newblock {\it \bibinfo{journal}{Proceedings of the IEEE}\/},  {\it
  \bibinfo{volume}{92}\/}, \bibinfo{pages}{423--438}.
  \DOIprefix\doi{10.1109/JPROC.2003.823142}.
%Type = Article
\bibitem[{Arasaratnam \& Haykin(2009)}]{arasaratnam_cubature_2009}
\bibinfo{author}{Arasaratnam, I.}, \& \bibinfo{author}{Haykin, S.}
  (\bibinfo{year}{2009}).
\newblock \bibinfo{title}{Cubature {Kalman} {Filters}}.
\newblock {\it \bibinfo{journal}{IEEE Transactions on Automatic Control}\/},
  {\it \bibinfo{volume}{54}\/}, \bibinfo{pages}{1254--1269}.
  \DOIprefix\doi{10.1109/TAC.2009.2019800}.
%Type = Article
\bibitem[{Aravkin et~al.(2017)Aravkin, Burke, Ljung, Lozano \&
  Pillonetto}]{aravkin_generalized_2017}
\bibinfo{author}{Aravkin, A.}, \bibinfo{author}{Burke, J.~V.},
  \bibinfo{author}{Ljung, L.}, \bibinfo{author}{Lozano, A.}, \&
  \bibinfo{author}{Pillonetto, G.} (\bibinfo{year}{2017}).
\newblock \bibinfo{title}{Generalized {Kalman} smoothing: {Modeling} and
  algorithms}.
\newblock {\it \bibinfo{journal}{Automatica}\/},  {\it \bibinfo{volume}{86}\/},
  \bibinfo{pages}{63--86}. \DOIprefix\doi{10.1016/j.automatica.2017.08.011}.
%Type = Inbook
\bibitem[{Aravkin et~al.(2014)Aravkin, Burke \&
  Pillonetto}]{aravkin_optimization_2014}
\bibinfo{author}{Aravkin, A.~Y.}, \bibinfo{author}{Burke, J.~V.}, \&
  \bibinfo{author}{Pillonetto, G.} (\bibinfo{year}{2014}).
\newblock \bibinfo{title}{Optimization {Viewpoint} on {Kalman} {Smoothing} with
  {Applications} to {Robust} and {Sparse} {Estimation}}.
\newblock In \bibinfo{editor}{A.~Y. Carmi}, \bibinfo{editor}{L.~Mihaylova}, \&
  \bibinfo{editor}{S.~J. Godsill} (Eds.), {\it \bibinfo{booktitle}{Compressed
  {Sensing} \& {Sparse} {Filtering}}\/} (pp. \bibinfo{pages}{237--280}).
\newblock \bibinfo{address}{Berlin, Heidelberg}: \bibinfo{publisher}{Springer
  Berlin Heidelberg}.
\newblock \DOIprefix\doi{10.1007/978-3-642-38398-4_8}.
%Type = Article
\bibitem[{{\AA}ström(1980)}]{astrom_maximum_1980}
\bibinfo{author}{{\AA}ström, K.~J.} (\bibinfo{year}{1980}).
\newblock \bibinfo{title}{Maximum likelihood and prediction error methods}.
\newblock {\it \bibinfo{journal}{Automatica}\/},  {\it \bibinfo{volume}{16}\/},
  \bibinfo{pages}{551--574}. \DOIprefix\doi{10.1016/0005-1098(80)90078-3}.
%Type = Article
\bibitem[{Barfoot et~al.(2020)Barfoot, Forbes \& Yoon}]{barfoot_exactly_2020}
\bibinfo{author}{Barfoot, T.~D.}, \bibinfo{author}{Forbes, J.~R.}, \&
  \bibinfo{author}{Yoon, D.~J.} (\bibinfo{year}{2020}).
\newblock \bibinfo{title}{Exactly sparse {Gaussian} variational inference with
  application to derivative-free batch nonlinear state estimation}.
\newblock {\it \bibinfo{journal}{The International Journal of Robotics
  Research}\/},  {\it \bibinfo{volume}{39}\/}, \bibinfo{pages}{1473--1502}.
  \DOIprefix\doi{10.1177/0278364920937608}.
%Type = Book
\bibitem[{Candy(2016)}]{candy_bayesian_2016}
\bibinfo{author}{Candy, J.~V.} (\bibinfo{year}{2016}).
\newblock {\it \bibinfo{title}{Bayesian Signal Processing: Classical, Modern,
  and Particle Filtering Methods}\/}.
\newblock \bibinfo{address}{Hoboken, NJ, USA}: \bibinfo{publisher}{John Wiley
  \& Sons, Inc.}
\newblock \DOIprefix\doi{10.1002/9781119125495}.
%Type = Article
\bibitem[{Courts et~al.(2021{\natexlab{a}})Courts, Hendriks, Wills, Schön \&
  Ninness}]{courts_variational_2021}
\bibinfo{author}{Courts, J.}, \bibinfo{author}{Hendriks, J.},
  \bibinfo{author}{Wills, A.}, \bibinfo{author}{Schön, T.~B.}, \&
  \bibinfo{author}{Ninness, B.} (\bibinfo{year}{2021}{\natexlab{a}}).
\newblock \bibinfo{title}{Variational {State} and {Parameter} {Estimation}}.
\newblock {\it \bibinfo{journal}{IFAC-PapersOnLine}\/},  {\it
  \bibinfo{volume}{54}\/}, \bibinfo{pages}{732--737}.
  \DOIprefix\doi{10.1016/j.ifacol.2021.08.448}.
%Type = Article
\bibitem[{Courts et~al.(2021{\natexlab{b}})Courts, Wills \&
  Schon}]{courts_gaussian_2021}
\bibinfo{author}{Courts, J.}, \bibinfo{author}{Wills, A.}, \&
  \bibinfo{author}{Schon, T.} (\bibinfo{year}{2021}{\natexlab{b}}).
\newblock \bibinfo{title}{Gaussian {Variational} {State} {Estimation} for
  {Nonlinear} {State}-{Space} {Models}}.
\newblock {\it \bibinfo{journal}{IEEE Transactions on Signal Processing}\/},
  {\it \bibinfo{volume}{69}\/}, \bibinfo{pages}{5979--5993}.
  \DOIprefix\doi{10.1109/TSP.2021.3122296}.
%Type = Article
\bibitem[{Courts et~al.(2023)Courts, Wills, Schön \&
  Ninness}]{courts_variational_2023}
\bibinfo{author}{Courts, J.}, \bibinfo{author}{Wills, A.~G.},
  \bibinfo{author}{Schön, T.~B.}, \& \bibinfo{author}{Ninness, B.}
  (\bibinfo{year}{2023}).
\newblock \bibinfo{title}{Variational system identification for nonlinear
  state-space models}.
\newblock {\it \bibinfo{journal}{Automatica}\/},  {\it
  \bibinfo{volume}{147}\/}, \bibinfo{pages}{110687}.
  \DOIprefix\doi{10.1016/j.automatica.2022.110687}.
%Type = Article
\bibitem[{Dellaert \& Kaess(2017)}]{dellaert_factor_2017}
\bibinfo{author}{Dellaert, F.}, \& \bibinfo{author}{Kaess, M.}
  (\bibinfo{year}{2017}).
\newblock \bibinfo{title}{Factor {Graphs} for {Robot} {Perception}}.
\newblock {\it \bibinfo{journal}{Foundations and Trends in Robotics}\/},  {\it
  \bibinfo{volume}{6}\/}, \bibinfo{pages}{1--139}.
  \DOIprefix\doi{10.1561/2300000043}.
%Type = Article
\bibitem[{Dutra et~al.(2014)Dutra, Teixeira \& Aguirre}]{dutra_maximum_2014}
\bibinfo{author}{Dutra, D.~A.}, \bibinfo{author}{Teixeira, B. O.~S.}, \&
  \bibinfo{author}{Aguirre, L.~A.} (\bibinfo{year}{2014}).
\newblock \bibinfo{title}{Maximum a posteriori state path estimation:
  {Discretization} limits and their interpretation}.
\newblock {\it \bibinfo{journal}{Automatica}\/},  {\it \bibinfo{volume}{50}\/},
  \bibinfo{pages}{1360--1368}.
  \DOIprefix\doi{10.1016/j.automatica.2014.03.003}.
%Type = Article
\bibitem[{Dutra(2020)}]{dutra_uncertainty_2020}
\bibinfo{author}{Dutra, D. A.~A.} (\bibinfo{year}{2020}).
\newblock \bibinfo{title}{Uncertainty estimation in equality-constrained {MAP}
  and maximum likelihood estimation with applications to system identification
  and state estimation}.
\newblock {\it \bibinfo{journal}{Automatica}\/},  {\it
  \bibinfo{volume}{116}\/}, \bibinfo{pages}{108935}.
  \DOIprefix\doi{10.1016/j.automatica.2020.108935}.
%Type = Article
\bibitem[{Dutra et~al.(2017)Dutra, Teixeira \& Aguirre}]{dutra_joint_2017}
\bibinfo{author}{Dutra, D. A.~A.}, \bibinfo{author}{Teixeira, B. O.~S.}, \&
  \bibinfo{author}{Aguirre, L.~A.} (\bibinfo{year}{2017}).
\newblock \bibinfo{title}{Joint maximum a posteriori state path and parameter
  estimation in stochastic differential equations}.
\newblock {\it \bibinfo{journal}{Automatica}\/},  {\it \bibinfo{volume}{81}\/},
  \bibinfo{pages}{403--408}. \DOIprefix\doi{10.1016/j.automatica.2017.03.035}.
%Type = Article
\bibitem[{Farahmand et~al.(2011)Farahmand, Giannakis \&
  Angelosante}]{farahmand_doubly_2011}
\bibinfo{author}{Farahmand, S.}, \bibinfo{author}{Giannakis, G.}, \&
  \bibinfo{author}{Angelosante, D.} (\bibinfo{year}{2011}).
\newblock \bibinfo{title}{Doubly {Robust} {Smoothing} of {Dynamical}
  {Processes} via {Outlier} {Sparsity} {Constraints}}.
\newblock {\it \bibinfo{journal}{IEEE Transactions on Signal Processing}\/},
  {\it \bibinfo{volume}{59}\/}, \bibinfo{pages}{4529--4543}.
  \DOIprefix\doi{10.1109/TSP.2011.2161300}.
%Type = Article
\bibitem[{Gedon et~al.(2021)Gedon, Wahlström, Schön \&
  Ljung}]{gedon_deep_2021}
\bibinfo{author}{Gedon, D.}, \bibinfo{author}{Wahlström, N.},
  \bibinfo{author}{Schön, T.~B.}, \& \bibinfo{author}{Ljung, L.}
  (\bibinfo{year}{2021}).
\newblock \bibinfo{title}{Deep {State} {Space} {Models} for {Nonlinear}
  {System} {Identification}}.
\newblock {\it \bibinfo{journal}{IFAC-PapersOnLine}\/},  {\it
  \bibinfo{volume}{54}\/}, \bibinfo{pages}{481--486}.
  \DOIprefix\doi{10.1016/j.ifacol.2021.08.406}.
%Type = Article
\bibitem[{Gibson \& Ninness(2005)}]{gibson_robust_2005}
\bibinfo{author}{Gibson, S.}, \& \bibinfo{author}{Ninness, B.}
  (\bibinfo{year}{2005}).
\newblock \bibinfo{title}{Robust maximum-likelihood estimation of multivariable
  dynamic systems}.
\newblock {\it \bibinfo{journal}{Automatica}\/},  {\it \bibinfo{volume}{41}\/},
  \bibinfo{pages}{1667--1682}.
  \DOIprefix\doi{10.1016/j.automatica.2005.05.008}.
%Type = Inproceedings
\bibitem[{Glasmachers et~al.(2010)Glasmachers, Schaul, Yi, Wierstra \&
  Schmidhuber}]{glasmachers_exponential_2010}
\bibinfo{author}{Glasmachers, T.}, \bibinfo{author}{Schaul, T.},
  \bibinfo{author}{Yi, S.}, \bibinfo{author}{Wierstra, D.}, \&
  \bibinfo{author}{Schmidhuber, J.} (\bibinfo{year}{2010}).
\newblock \bibinfo{title}{Exponential natural evolution strategies}.
\newblock In {\it \bibinfo{booktitle}{Proceedings of the 12th annual conference
  on {Genetic} and evolutionary computation}\/} (pp.
  \bibinfo{pages}{393--400}).
\newblock \bibinfo{publisher}{ACM}.
\newblock \DOIprefix\doi{10.1145/1830483.1830557}.
%Type = Inproceedings
\bibitem[{Grauer \& Morelli(2015)}]{grauer_new_2015}
\bibinfo{author}{Grauer, J.}, \& \bibinfo{author}{Morelli, E.}
  (\bibinfo{year}{2015}).
\newblock \bibinfo{title}{A {New} {Formulation} of the {Filter}-{Error}
  {Method} for {Aerodynamic} {Parameter} {Estimation} in {Turbulence}}.
\newblock In {\it \bibinfo{booktitle}{{AIAA} {Atmospheric} {Flight} {Mechanics}
  {Conference}}\/}.
\newblock \DOIprefix\doi{10.2514/6.2015-2704}.
%Type = Book
\bibitem[{Jategaonkar(2015)}]{jategaonkar_flight_2015}
\bibinfo{author}{Jategaonkar, R.} (\bibinfo{year}{2015}).
\newblock {\it \bibinfo{title}{Flight {Vehicle} {System} {Identification}}\/}.
\newblock (\bibinfo{edition}{Second ed.} ed.).
\newblock \bibinfo{address}{Reston, VA}: \bibinfo{publisher}{AIAA}.
\newblock \DOIprefix\doi{10.2514/4.102790}.
%Type = Book
\bibitem[{Kailath et~al.(2000)Kailath, Sayed \& Hassibi}]{kailath_linear_2000}
\bibinfo{author}{Kailath, T.}, \bibinfo{author}{Sayed, A.~H.}, \&
  \bibinfo{author}{Hassibi, B.} (\bibinfo{year}{2000}).
\newblock {\it \bibinfo{title}{Linear estimation}\/}.
\newblock Prentice {Hall} information and system sciences series.
\newblock \bibinfo{address}{Upper Saddle River, N.J}:
  \bibinfo{publisher}{Prentice Hall}.
%Type = Article
\bibitem[{Karimi \& McAuley(2013)}]{karimi_approximate_2013}
\bibinfo{author}{Karimi, H.}, \& \bibinfo{author}{McAuley, K.~B.}
  (\bibinfo{year}{2013}).
\newblock \bibinfo{title}{An {Approximate} {Expectation} {Maximization}
  {Algorithm} for {Estimating} {Parameters}, {Noise} {Variances}, and
  {Stochastic} {Disturbance} {Intensities} in {Nonlinear} {Dynamic} {Models}}.
\newblock {\it \bibinfo{journal}{Industrial \& Engineering Chemistry
  Research}\/},  {\it \bibinfo{volume}{52}\/}, \bibinfo{pages}{18303--18323}.
  \DOIprefix\doi{10.1021/ie4023989}.
%Type = Article
\bibitem[{Karimi \& McAuley(2014)}]{karimi_maximum-likelihood_2014}
\bibinfo{author}{Karimi, H.}, \& \bibinfo{author}{McAuley, K.~B.}
  (\bibinfo{year}{2014}).
\newblock \bibinfo{title}{A maximum-likelihood method for estimating
  parameters, stochastic disturbance intensities and measurement noise
  variances in nonlinear dynamic models with process disturbances}.
\newblock {\it \bibinfo{journal}{Computers \& Chemical Engineering}\/},  {\it
  \bibinfo{volume}{67}\/}, \bibinfo{pages}{178--198}.
  \DOIprefix\doi{10.1016/j.compchemeng.2014.04.007}.
%Type = Article
\bibitem[{Karimi \& McAuley(2018)}]{karimi_bayesian_2018}
\bibinfo{author}{Karimi, H.}, \& \bibinfo{author}{McAuley, K.~B.}
  (\bibinfo{year}{2018}).
\newblock \bibinfo{title}{Bayesian {Objective} {Functions} for {Estimating}
  {Parameters} in {Nonlinear} {Stochastic} {Differential} {Equation} {Models}
  with {Limited} {Data}}.
\newblock {\it \bibinfo{journal}{Industrial \& Engineering Chemistry
  Research}\/},  {\it \bibinfo{volume}{57}\/}, \bibinfo{pages}{8946--8961}.
  \DOIprefix\doi{10.1021/acs.iecr.8b00293}.
%Type = Inproceedings
\bibitem[{Kingma \& Ba(2015)}]{kingma_adam_2015}
\bibinfo{author}{Kingma, D.~P.}, \& \bibinfo{author}{Ba, J.}
  (\bibinfo{year}{2015}).
\newblock \bibinfo{title}{Adam: {A} {Method} for {Stochastic} {Optimization}}.
\newblock In {\it \bibinfo{booktitle}{Proceedings of the 3rd {International}
  {Conference} for {Learning} {Representations}}\/}.
\newblock \bibinfo{address}{San Diego, CA, USA}.
%Type = Book
\bibitem[{Kloeden \& Platen(1992)}]{kloeden_numerical_1992}
\bibinfo{author}{Kloeden, P.~E.}, \& \bibinfo{author}{Platen, E.}
  (\bibinfo{year}{1992}).
\newblock {\it \bibinfo{title}{Numerical {Solution} of {Stochastic}
  {Differential} {Equations}}\/}.
\newblock Number~\bibinfo{number}{23} in \bibinfo{series}{Applications of
  {Mathematics}} (\bibinfo{edition}{corrected third printing} ed.).
\newblock \bibinfo{publisher}{Springer}.
%Type = Article
\bibitem[{Kokkala et~al.(2016)Kokkala, Solin \&
  Särkkä}]{kokkala_sigma-point_2016}
\bibinfo{author}{Kokkala, J.}, \bibinfo{author}{Solin, A.}, \&
  \bibinfo{author}{Särkkä, S.} (\bibinfo{year}{2016}).
\newblock \bibinfo{title}{Sigma-{Point} {Filtering} and {Smoothing} {Based}
  {Parameter} {Estimation} in {Nonlinear} {Dynamic} {Systems}}.
\newblock {\it \bibinfo{journal}{Journal of Advances in Information Fusion}\/},
   {\it \bibinfo{volume}{11}\/}, \bibinfo{pages}{15--30}.
%Type = Article
\bibitem[{Kristensen et~al.(2004)Kristensen, Madsen \&
  Jørgensen}]{kristensen_parameter_2004}
\bibinfo{author}{Kristensen, N.}, \bibinfo{author}{Madsen, H.}, \&
  \bibinfo{author}{Jørgensen, S.} (\bibinfo{year}{2004}).
\newblock \bibinfo{title}{Parameter estimation in stochastic grey-box models}.
\newblock {\it \bibinfo{journal}{Automatica}\/},  {\it \bibinfo{volume}{40}\/},
  \bibinfo{pages}{225--237}. \DOIprefix\doi{10.1016/j.automatica.2003.10.001}.
%Type = Book
\bibitem[{Ljung(1999)}]{ljung_system_1999}
\bibinfo{author}{Ljung, L.} (\bibinfo{year}{1999}).
\newblock {\it \bibinfo{title}{System identification: theory for the user}\/}.
\newblock Prentice {Hall} information and system sciences series
  (\bibinfo{edition}{2nd} ed.).
\newblock \bibinfo{address}{Upper Saddle River, NJ}:
  \bibinfo{publisher}{Prentice Hall PTR}.
%Type = Book
\bibitem[{MacKay(2019)}]{mackay_information_2019}
\bibinfo{author}{MacKay, D. J.~C.} (\bibinfo{year}{2019}).
\newblock {\it \bibinfo{title}{Information theory, inference, and learning
  algorithms}\/}.
\newblock (\bibinfo{edition}{fourth printing} ed.).
\newblock \bibinfo{address}{Cambridge}: \bibinfo{publisher}{Cambridge
  University Press}.
%Type = Book
\bibitem[{Magnus \& Neudecker(2019)}]{magnus_matrix_2019}
\bibinfo{author}{Magnus, J.~R.}, \& \bibinfo{author}{Neudecker, H.}
  (\bibinfo{year}{2019}).
\newblock {\it \bibinfo{title}{Matrix {Differential} {Calculus} with
  {Applications} in {Statistics} and {Econometrics}}\/}.
\newblock Wiley {Series} in {Probability} and {Statistics}.
\newblock \bibinfo{publisher}{Wiley}.
\newblock \DOIprefix\doi{10.1002/9781119541219}.
%Type = Article
\bibitem[{Maine \& Iliff(1981)}]{maine_formulation_1981}
\bibinfo{author}{Maine, R.~E.}, \& \bibinfo{author}{Iliff, K.~W.}
  (\bibinfo{year}{1981}).
\newblock \bibinfo{title}{Formulation and {Implementation} of a {Practical}
  {Algorithm} for {Parameter} {Estimation} with {Process} and {Measurement}
  {Noise}}.
\newblock {\it \bibinfo{journal}{SIAM Journal on Applied Mathematics}\/},  {\it
  \bibinfo{volume}{41}\/}, \bibinfo{pages}{558--579}.
  \DOIprefix\doi{10.1137/0141045}.
%Type = Article
\bibitem[{McKelvey et~al.(2004)McKelvey, Helmersson \&
  Ribarits}]{mckelvey_data_2004}
\bibinfo{author}{McKelvey, T.}, \bibinfo{author}{Helmersson, A.}, \&
  \bibinfo{author}{Ribarits, T.} (\bibinfo{year}{2004}).
\newblock \bibinfo{title}{Data driven local coordinates for multivariable
  linear systems and their application to system identification}.
\newblock {\it \bibinfo{journal}{Automatica}\/},  {\it \bibinfo{volume}{40}\/},
  \bibinfo{pages}{1629--1635}.
  \DOIprefix\doi{10.1016/j.automatica.2004.04.015}.
%Type = Article
\bibitem[{Opper \& Archambeau(2009)}]{opper_variational_2009}
\bibinfo{author}{Opper, M.}, \& \bibinfo{author}{Archambeau, C.}
  (\bibinfo{year}{2009}).
\newblock \bibinfo{title}{The {Variational} {Gaussian} {Approximation}
  {Revisited}}.
\newblock {\it \bibinfo{journal}{Neural Computation}\/},  {\it
  \bibinfo{volume}{21}\/}, \bibinfo{pages}{786--792}.
  \DOIprefix\doi{10.1162/neco.2008.08-07-592}.
%Type = Inproceedings
\bibitem[{Ribeiro et~al.(2020{\natexlab{a}})Ribeiro, Tiels, Aguirre \&
  Sch\"on}]{ribeiro_beyond_2020}
\bibinfo{author}{Ribeiro, A.~H.}, \bibinfo{author}{Tiels, K.},
  \bibinfo{author}{Aguirre, L.~A.}, \& \bibinfo{author}{Sch\"on, T.~B.}
  (\bibinfo{year}{2020}{\natexlab{a}}).
\newblock \bibinfo{title}{Beyond exploding and vanishing gradients: analysing
  rnn training using attractors and smoothness}.
\newblock In {\it \bibinfo{booktitle}{Proceedings of the 23rd International
  Conference on Artificial Intelligence and Statistics (AISTATS).}\/}.
\newblock \bibinfo{publisher}{PMLR} volume \bibinfo{volume}{108}.
%Type = Article
\bibitem[{Ribeiro et~al.(2020{\natexlab{b}})Ribeiro, Tiels, Umenberger, Schön
  \& Aguirre}]{ribeiro_smoothness_2020}
\bibinfo{author}{Ribeiro, A.~H.}, \bibinfo{author}{Tiels, K.},
  \bibinfo{author}{Umenberger, J.}, \bibinfo{author}{Schön, T.~B.}, \&
  \bibinfo{author}{Aguirre, L.~A.} (\bibinfo{year}{2020}{\natexlab{b}}).
\newblock \bibinfo{title}{On the smoothness of nonlinear system
  identification}.
\newblock {\it \bibinfo{journal}{Automatica}\/},  {\it
  \bibinfo{volume}{121}\/}, \bibinfo{pages}{109158}.
  \DOIprefix\doi{10.1016/j.automatica.2020.109158}.
%Type = Inproceedings
\bibitem[{Schmidt et~al.(2021)Schmidt, Schneider \&
  Hennig}]{schmidt_descending_2021}
\bibinfo{author}{Schmidt, R.~M.}, \bibinfo{author}{Schneider, F.}, \&
  \bibinfo{author}{Hennig, P.} (\bibinfo{year}{2021}).
\newblock \bibinfo{title}{Descending through a {Crowded} {Valley} -
  {Benchmarking} {Deep} {Learning} {Optimizers}}.
\newblock In {\it \bibinfo{booktitle}{Proceedings of the 38th {International}
  {Conference} on {Machine} {Learning}}\/} (pp. \bibinfo{pages}{9367--9376}).
\newblock \bibinfo{publisher}{PMLR} volume \bibinfo{volume}{139} of {\it
  \bibinfo{series}{Proceedings of {Machine} {Learning} {Research}}\/}.
%Type = Article
\bibitem[{Schön et~al.(2011)Schön, Wills \& Ninness}]{schon_system_2011}
\bibinfo{author}{Schön, T.}, \bibinfo{author}{Wills, A.}, \&
  \bibinfo{author}{Ninness, B.} (\bibinfo{year}{2011}).
\newblock \bibinfo{title}{System identification of nonlinear state-space
  models}.
\newblock {\it \bibinfo{journal}{Automatica}\/},  {\it \bibinfo{volume}{47}\/},
  \bibinfo{pages}{39--49}. \DOIprefix\doi{10.1016/j.automatica.2010.10.013}.
%Type = Book
\bibitem[{Särkkä(2013)}]{sarkka_bayesian_2013}
\bibinfo{author}{Särkkä, S.} (\bibinfo{year}{2013}).
\newblock {\it \bibinfo{title}{Bayesian Filtering and Smoothing}\/}.
\newblock \bibinfo{address}{Cambridge}: \bibinfo{publisher}{Cambridge
  University Press}.
\newblock \DOIprefix\doi{10.1017/CBO9781139344203}.
%Type = Article
\bibitem[{Tzikas et~al.(2008)Tzikas, Likas \&
  Galatsanos}]{tzikas_variational_2008}
\bibinfo{author}{Tzikas, D.~G.}, \bibinfo{author}{Likas, A.~C.}, \&
  \bibinfo{author}{Galatsanos, N.~P.} (\bibinfo{year}{2008}).
\newblock \bibinfo{title}{The variational approximation for {Bayesian}
  inference}.
\newblock {\it \bibinfo{journal}{IEEE Signal Processing Magazine}\/},  {\it
  \bibinfo{volume}{25}\/}, \bibinfo{pages}{131--146}.
  \DOIprefix\doi{10.1109/MSP.2008.929620}.
%Type = Article
\bibitem[{Virtanen et~al.(2020)Virtanen, Gommers, Oliphant
  et~al.}]{virtanen_scipy_2020}
\bibinfo{author}{Virtanen, P.}, \bibinfo{author}{Gommers, R.},
  \bibinfo{author}{Oliphant, T.~E.} et~al. (\bibinfo{year}{2020}).
\newblock \bibinfo{title}{{SciPy} 1.0: fundamental algorithms for scientific
  computing in {Python}}.
\newblock {\it \bibinfo{journal}{Nature Methods}\/},  {\it
  \bibinfo{volume}{17}\/}, \bibinfo{pages}{261--272}.
  \DOIprefix\doi{10.1038/s41592-019-0686-2}.
%Type = Article
\bibitem[{Wills \& Ninness(2008)}]{wills_gradient-based_2008}
\bibinfo{author}{Wills, A.}, \& \bibinfo{author}{Ninness, B.}
  (\bibinfo{year}{2008}).
\newblock \bibinfo{title}{On {Gradient}-{Based} {Search} for {Multivariable}
  {System} {Estimates}}.
\newblock {\it \bibinfo{journal}{IEEE Transactions on Automatic Control}\/},
  {\it \bibinfo{volume}{53}\/}, \bibinfo{pages}{298--306}.
  \DOIprefix\doi{10.1109/TAC.2007.914953}.
%Type = Article
\bibitem[{Wills \& Schön(2021)}]{wills_stochastic_2021}
\bibinfo{author}{Wills, A.~G.}, \& \bibinfo{author}{Schön, T.~B.}
  (\bibinfo{year}{2021}).
\newblock \bibinfo{title}{Stochastic quasi-{Newton} with line-search
  regularisation}.
\newblock {\it \bibinfo{journal}{Automatica}\/},  {\it
  \bibinfo{volume}{127}\/}, \bibinfo{pages}{109503}.
  \DOIprefix\doi{10.1016/j.automatica.2021.109503}.
%Type = Article
\bibitem[{Zhang et~al.(2019)Zhang, Butepage, Kjellstrom \&
  Mandt}]{zhang_advances_2019}
\bibinfo{author}{Zhang, C.}, \bibinfo{author}{Butepage, J.},
  \bibinfo{author}{Kjellstrom, H.}, \& \bibinfo{author}{Mandt, S.}
  (\bibinfo{year}{2019}).
\newblock \bibinfo{title}{Advances in {Variational} {Inference}}.
\newblock {\it \bibinfo{journal}{IEEE Transactions on Pattern Analysis and
  Machine Intelligence}\/},  {\it \bibinfo{volume}{41}\/},
  \bibinfo{pages}{2008--2026}. \DOIprefix\doi{10.1109/TPAMI.2018.2889774}.

\end{thebibliography}

\end{document}